\documentclass[12pt]{article}

    \usepackage[utf8]{inputenc}
    \usepackage[
        twoside,
        top=1in,
        bottom=0.75in,
        inner=0.75in,
        outer=0.75in
    ]{geometry}
    \usepackage[square, numbers]{natbib}
    \usepackage{color}

    \usepackage{amsfonts,amsmath,amssymb}
    \usepackage{titling}

    \usepackage{graphicx}
    \usepackage{hyperref}
    
    \usepackage{caption}
    \usepackage{subcaption}

    \usepackage[export]{adjustbox}

    \usepackage{multirow}

    \usepackage{mathtools}

    \usepackage{booktabs} 
    \usepackage{array}

    \usepackage{authblk}
    
    \newcommand{\avgneutralspd}{\mu_v}
    \newcommand{\meanidealangular}{\mu_\omega}
    \newcommand{\varianceidealangular}{\sigma_\omega}
    
    \newcommand{\tblwidth}{\linewidth}
    \newcolumntype{L}{l}
    \newcolumntype{R}{r}
    \newcolumntype{C}{c}

    \title{Microscopic modeling of attention-based movement behaviors}
    \author[1,*]{Danrui Li}
    \author[1]{Mathew Schwartz}
    \author[1]{Samuel S. Sohn}
    \author[2]{Sejong Yoon}
    \author[1]{Vladimir Pavlovic}
    \author[1]{Mubbasir Kapadia}
    \affil[1]{\small Rutgers University, 08854, New Jersey, USA}
    \affil[2]{The College of New Jersey, 08628, New Jersey, USA}
    \affil[*]{Corresponding author: \texttt{danrui.li@rutgers.edu}}

    \date{}

\begin{document}
    
\markboth{Danrui Li, Mathew Schwartz, Samuel S. Sohn et al.}{\thetitle}

\maketitle

\begin{abstract}
For transportation hubs, leveraging pedestrian flows for commercial activities presents an effective strategy for funding maintenance and infrastructure improvements. However, this introduces new challenges, as consumer behaviors can disrupt pedestrian flow and efficiency. To optimize both retail potential and pedestrian efficiency, careful strategic planning in store layout and facility dimensions was done by expert judgement due to the complexity in pedestrian dynamics in the retail areas of transportation hubs.
This paper introduces an attention-based movement model to simulate these dynamics. By simulating retail potential of an area through the duration of visual attention it receives, and pedestrian efficiency via speed loss in pedestrian walking behaviors, the study further explores how design features can influence the retail potential and pedestrian efficiency in a bi-directional corridor inside a transportation hub. Project webpage: \href{https://danruili.github.io/AttentionMove/}{https://danruili.github.io/AttentionMove/} 

\end{abstract}

\section{Introduction}\label{sec:intro}

Pedestrian dynamics within public spaces have garnered significant attention in recent years due to their implications for the planning of transportation facilities, commercial centers, and other urban infrastructures \citep{Hoogendoorn2004}. While pedestrian safety has been the focus for a long time, recent demands call for explorations about the pedestrian dynamics in retail areas within high-density transportation hubs. As passenger growth increases the expenditure of transportation facility operations, leveraging such pedestrian flows to expand commercial activities becomes a practical way for the management to fund maintenance and asset improvements\citep{baron_designing_2019}.

However, while increases in pedestrian flows provide financial benefits, research suggests consuming behaviors may introduce slow-down effects or turbulence to pedestrian flows \citep{zacharias_pedestrian_2021}. These conflicting outcomes create a challenge when incorporating retail areas into high-density transportation hubs. In order to maintain the efficiency requirements of transportation services, architectural designers must strike a balance between retail potential and pedestrian efficiency through strategic store layout and facility dimensions \citep{hanseler}. This process is currently done through intuition and expert judgement, in part due to the difficulties in simulating the possible outcomes given modifications to the design and layout.

One of the difficulties is the diversity of pedestrian interactions with environmental objects, which limits the applications of previous studies. Most prior works follow an assumption that pedestrians tend to steer towards environmental objects when they are attracted to them \citep{kwak_collective_2013, zhou_modeling_2022}. However, \cite{Li2022} shows that such a model does not impose the dominant impact on the slowing-down effects in transportation hub retail areas. Instead, pedestrians who slow down to gaze at stores without changing directions actually provide the major contributions \citep{Li2022}. Such a behavior, which we refer to as \textit{attention-based movement}, differ from endogenous distracted movement in \cite{kremer_modelling_2021} because in our context agents will interact with exogenous environmental objects rather than something they hold. Instead, our work is related to the attracted movement behaviors described in \cite{weili_wang_modeling_2014}. Since the modelling of attention-based movement remains largely unexplored, in this study we focus on this subgroup to pave the way to the design optimization goal above.

Another challenge is to find metrics for pedestrian efficiency and retail potential, then model them simultaneously. Retail potential, often referred to as the probability of purchasing a product/item, is unable to directly inspect in our study. Because we focus on attention-based movements where pedestrians do not enter the store to purchase. However, retail potential is found to be correlated with the time of looking at the products (the duration of visual attention) \citep{gidlof_looking_2017}. Such influence becomes even stronger in rapid decisions \citep{milosavljevic_relative_2012}, which are often the cases in transportation hubs. Therefore, we consider the average visual attention that pedestrians pay to a retail area as a reasonable measure for the retail potential of it. It can also be seen as the visual saliency of retail areas from an exocentric perspective \citep{kremer_modelling_2021}. Pedestrian efficiency, on the other hand, can be measured by the speed loss compared to pedestrians' desired walking speed \citep{helbing_freezing_2000, kwak_collective_2013}. Therefore, we need to simulate not only pedestrians' visual attention but also their changes in walking behaviors when they are interacting with environmental objects.

Therefore, in this paper, we propose a novel pedestrian model referred to as \textit{attention-based movement behaviors}. Our model is a combination of visual attention and locomotion, designed for simulating attention within retail transportation hubs. Section \ref{sec:relatedwork} provides an overview of related work. Section \ref{sec:3} describes the dataset, models, and parameter settings. Section \ref{sec:4} compares its simulation results to the real-world dataset and prior work. Then, our model will be applied to a sensitivity analysis with regard to the architectural design features in Section \ref{sec:5} to demonstrate its capability in design optimization. Finally, the paper ends with a discussion and conclusion in section \ref{sec:6}. 

\section{Related Work}\label{sec:relatedwork}

\subsection{Visual attention}

In this paper, the visual attention of a pedestrian is a status when the pedestrian moves their eyes to a location and the focus of attention is in accordance with eye movements, referred to as \textit{spatial overt attention} (compared to the attention that does not involve eye movements)~\citep{carrasco_visual_2011}.  Visual attention can be triggered by not only internal factors (e.g., pedestrians' preferences)~\citep{gidlof_looking_2017}, but also external factors such as the physical feature (color, luminance, etc.) of environmental objects~\citep{saunders_curious_2004}, the presence of surrounding pedestrians~\citep{gallup_visual_2012}, and visual coverage (one of the most explored aspects) \citep{xie_signage_2007, nassar_sign_2011, chen_attention_2011, wang_microscopic_2014, kremer_modelling_2021, zhou_modeling_2022}.

Visual coverage is the state of environmental objects covering pedestrians' visual perception fields. It describes not only the area they cover but also the location (i.e., the object is on the center of our retinal image or the fringe of that). In previous studies, visual coverage has been represented in different ways. \cite{xie_signage_2007} explored the boundary of visual perception fields (referred to as visual catchment area) using angular separation of environmental objects. \cite{chen_attention_2011} and \cite{wang_microscopic_2014} used distance and proportion of visual coverage, while \cite{nassar_sign_2011} used a piece-wise scaling term on the proportion of visual coverage and multiplied it by distance. \cite{zhou_modeling_2022} and \cite{kremer_watch_2020} employed observation angles and distances. Some of their representations of visual coverage are illustrated in Fig. \ref{fig:input}. We found that not all of these prior works can fully describe the area and the location of visual coverage at the same time (summarized in Table~\ref{tab:model_theo_compare}).

Additionally, pedestrians continuously shift their visual attention among various objects in retail areas, resulting in a series of attention initiation and termination events. This differs from evacuation scenarios where we care about how people initiate their attention to a sign rather than how they terminate it. In our context, we study the attention transitions in both directions (i.e. the initiation and the termination of attention). Furthermore, the visual attention model should be calibrated by real-world data, and must easily fit to data. Importantly, this model must have fast execution times due to the large number of agents (typically hundreds) involved in transportation hub simulations.

Our work is compared with prior literature in Table~\ref{tab:model_theo_compare} across: (1) Visual coverage representation in area and location (2) State transitions (3) Performant execution (4) Limitations specific to the models. Our work addresses gaps in prior works required for evaluating retail in transportation hubs. Specifically, we will build a visual attention model that can fully describe visual coverage, predict attention transitions in both directions calibrated by real-world data, and run fast.

\begin{table}[h]
    \caption{Visual attention module functionality comparison among previous work and ours.}
    \label{tab:model_theo_compare}
    \begin{tabular*}{\tblwidth}{@{}LLLLLL@{}}
    \toprule
    \multirow{2}{4em}{Prior work} & \multicolumn{2}{@{}L@{}}{Visual coverage representation}  & \multirow{2}{4em}{All state transitions} & \multirow{2}{4em}{Fast running} & \multirow{2}{4em}{Other limitations}\\ \cmidrule(r){2-3}
    
     & Area & Location & & & \\
    \midrule
         
     \cite{xie_signage_2007} & \checkmark & - & \checkmark & \checkmark & Not considering pedestrian orientations\\

     \cite{nassar_sign_2011} & \checkmark & partially & - & \checkmark & Not calibrated by real data \\ 

     \cite{wang_microscopic_2014} & \checkmark & partially & - & \checkmark & Not calibrated by real data\\ 

     \cite{kremer_watch_2020} & \checkmark & \checkmark & - & - & Not calibrated by real data\\

     \cite{zhou_modeling_2022} & partially & \checkmark & \checkmark & \checkmark &  Invariant to attraction sizes\\

     Ours & \checkmark &\checkmark &\checkmark &\checkmark &\\
    \bottomrule
    \end{tabular*}
    
\end{table}

\subsection{Locomotion in retail areas}

Locomotion models describe how pedestrians move in space at a resolution approximately of the human body. In the narrowest applications, they are used in simple environments without consideration for visual attention. The most prolific of which is the Social Force Model (SFM)~\citep{helbing1995social} and its subsequent variations~\citep{johansson_specification, karamouzas_predictive_2009,zanlungo_microscopic_2012}. Generally, the SFM determines pedestrian positions using their second derivatives that are summed by various hand-crafted forces. These forces represent the influences from walking goals, other pedestrians, the environment, etc.

Due to the popularity of the SFM, many previous studies that consider visual attention have built their models upon it. In \cite{kwak_collective_2013} and \cite{zhou_modeling_2022}, additional forces pointing to environmental objects were implemented to reflect the trend of pedestrians walking towards them. Nonetheless, as mentioned in \S \ref{sec:intro}, such a model design differs from the nature of attention-based movements, where pedestrians maintain their walking direction, but simultaneously draw attention to environmental objects. Such a model is therefore not considered in this paper.

\cite{weili_wang_modeling_2014} modeled attention-based movements by simply reducing walking speed to a fixed value when pedestrians draw their visual attention to environmental objects. But such a design is not supported by others' empirical findings, where pedestrians with visual attention showed varying speed, such as a reduction in walking speed as they approached the environmental objects~\citep{Li2022}.

While there is little literature about the varying walking speed when visual attention is drawn, some cognitive science experiments may provide insights. \cite{warren_behavioral_2008} claimed that pedestrian locomotion is guided by optical flow. Specifically, pedestrians try to maintain the angular position of their walking goals in their field of vision. We use this research as a reference point to explore similar rules for attention-based movements.

Therefore, in this paper, we will build a locomotion model for attention-based movements upon the SFM, and reveal how visual attention affects locomotion by the relative angular speed of the environmental objects.

\subsection{Pedestrian-simulation-based Design Evaluation}

Evaluating architectural design by crowd simulation has been a long-time practice~\citep{Hoogendoorn2004, feng_crowd-driven_2016, hu_predicting_2020}. However, the types of models vary in three key aspects. First, while many studies focus on egress behaviors in emergency scenarios~\citep{hu_predicting_2020}, we consider normal daily scenarios~\citep{feng_crowd-driven_2016} in which our evaluation metrics shift from safety (such as pedestrian density) to efficiency and retail potential. 

Second, compared to some prior works that evaluate the design on a large scale with a focus on the graphical configurations of entire floor plans~\citep{borgers_model_1986,feng_crowd-driven_2016}, we adopt a micro-scale perspective focusing on the dimensions of environmental objects, similar to ~\citep{shukla_genetically_2009,hu_predicting_2020}. This is because, for the majority of transportation hubs, their floor plans are dominated by their main functions, leaving little configurational flexibility to retail spaces.

Finally, previous studies vary in the study of independent variables. 
While some researchers aim to find optimized sets of design variables either by searching through continuous parameter space (summarized in \cite{nguyen_review_2014}) or by comparing several designs that are drafted by experts \citep{pantano_enhancing_2021,zhou_modeling_2022}, we want to get a more detailed understanding of the impacts of several variables by local sensitivity analysis or factorial designs, as in the work \citep{zhang_optimal_2017}. 

\section{Method}
\label{sec:3}

\begin{figure}[ht]
\centering
\includegraphics[width=\textwidth]{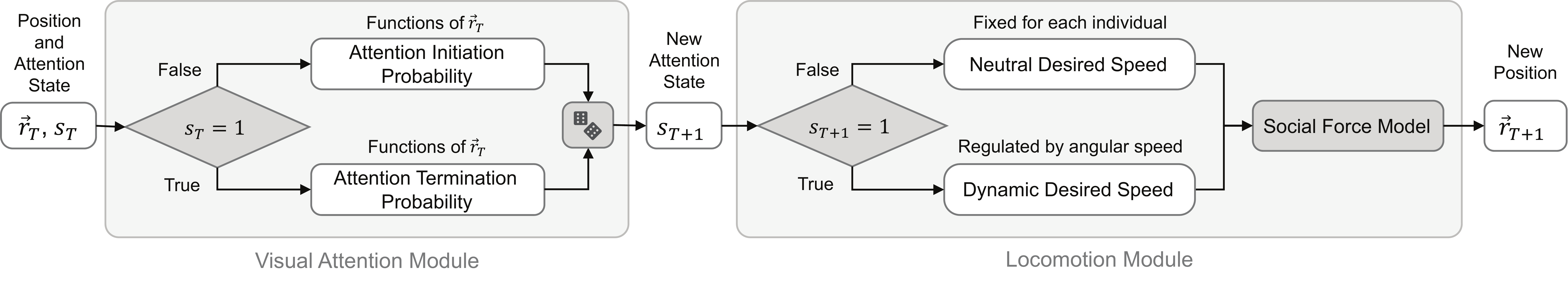}
\caption{
\textbf{Model Architecture}: The simulation framework is a loop composed of two modules: visual attention and locomotion. The loop begins with a position $\vec{r}_{T}$ and a visual attention state $s_T$ at time step $T$. Based on them, the probability of initiating or terminating the attention is calculated. From the probability, we get the attention state at the next time step $s_{T+1}$. Then the new attention state determines the desired speed of a pedestrian, which is fed into a Social Force Model to get a new position $\vec{r}_{T+1}$.}

\label{fig:framework}
\end{figure}

The attention transitions and walking speed variations in attention-based movement behaviors are jointly simulated as follows. For each virtual pedestrian, the simulation framework is a loop consisting of a visual attention module and a locomotion module (see Fig.~\ref{fig:framework}), which produce attention states and positions respectively at every time step. We define the attention state $s_T$ as a binary value, where $s_T=0$ means the pedestrian's attention is not drawn to the object and $s_T=1$ means the contrary.

The loop begins with the visual attention module (\S \ref{sec: visual}), where we derive the new visual attention state $s_{T+1}$ by calculating the probability of state transition from its previous state $s_{T}$. In other words, if the visual attention is not initiated ($s_T = 0$), we calculate the probability of initiating it. Otherwise, we calculate the probability of terminating it. Both probabilities are functions of position  $\vec{r}_{T}$.

In the locomotion module (\S \ref{locomotion}), we first calculate the desired walking speed of the pedestrian based on the visual attention state. If the pedestrian is not drawing attention to the environmental object ($s_{T+1}=0$), we set the desired walking speed to a neutral value, which is fixed for each pedestrian. Otherwise, the desired walking speed is determined by the angular speed to the environmental object (detailed in \S \ref{sec: angular}). Finally, the desired walking speed is used as an input of a Social Force Model, where the new position $\vec{r}_{T+1}$ is calculated (see \S \ref{sec: sfm integration}).

\subsection{Data Collection}

\begin{figure}[ht]
\centering
\includegraphics[width=8cm]{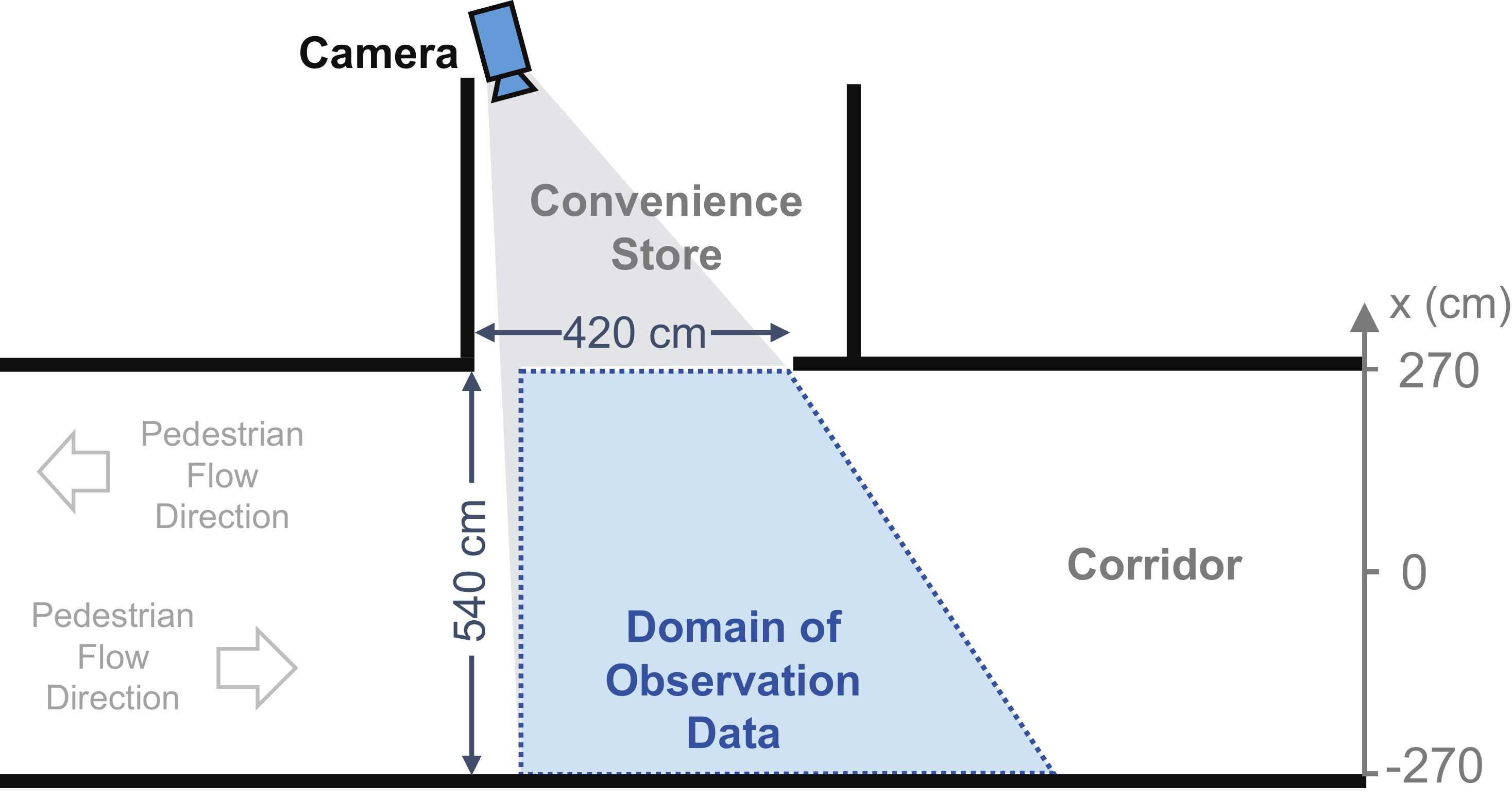}
\caption{\textbf{Field observation site and the camera installation}. The study area covers the front of a convenience store. And the origin of the x-axis is set to the middle point of the corridor section line. }
\label{fig:install}
\end{figure}

Our model is fit to empirical data by~\cite{position}. For clarity, we briefly describe the data collection process. Data was collected from an underground bi-directional corridor at a metro station. A solitary convenience store is situated within the corridor, devoid of additional stores, obstacles, or environmental objects that may get pedestrians' visual attention.

A camera was positioned at the store entrance, unobtrusively recording 45-minute video segments during off-peak daytime hours. After removing 97 pedestrians who went into the store, we captured the movements of 1,153 pedestrians. Due to the prevalence of pedestrian occlusions, pedestrian positions were manually labeled every 15 frames (0.5 seconds), drawing upon methodologies employed in prior research~\citep{gallup_visual_2012}. Then pedestrian trajectories were smoothed by cubic interpolation.

In this paper, a pedestrian's head turning towards the store is treated as a surrogate of their visual attention to it. The dependent variable -- the visual attention state to the store -- was manually labeled from video recordings every 0.5s as:

\begin{equation}
s_T = \begin{cases*}
  1, & if pedestrian turns head towards store,\\
  0,                    & otherwise.
\end{cases*}
\end{equation}

\subsection{Visual Attention Module} \label{sec: visual}

In this section, we firstly explain why we derive the visual attention state $s_{T+1}$ by calculating the probability of attention state transition from its previous state $s_{T}$ in \S \ref{sec: markov}. Then, we show how the position $\vec{r}_T$ is converted to two variables that fully describe the visual coverage information, and how the variables constitute functions that represent the probabilities of attention state transitions in \S \ref{sec: inputs}.

\subsubsection{Markov Chain structure} \label{sec: markov}

Predicting attention states over time can be treated as a sequence prediction task, where some previous studies~\citep{zhou_modeling_2022, wang_microscopic_2014} do not consider temporal dependency between prediction results. In those cases, the likelihood of a given sequence was formulated as $p(s_{1:T+1}) =p(s_{1})p(s_2) \cdots p(s_{T})p(s_T+1)$. While such models can successfully estimate the probability of attention status at time $T$, they fail to simulate the duration of behaviors as Markov chain does~\citep{haldi_interactions_2009}. 
To overcome the limitation, a first-order Markov chain is applied in this paper as: 
$p(s_{1:T+1}) =p(s_{1})p(s_2 | s_1) \cdots p(s_{T} | s_{T-1})p(s_{T+1} | s_{T})$,
where we predict attention state sequences by modeling the probabilities of the state transitions. The transition probability is only conditioned on the previous state. 

When fitting this model, we only need to model two probabilities: the probability of initiating visual attention when a pedestrian draws no attention to the object $p(s_{T+1} = 1 | s_{T} = 0)$, and that of the opposite event $p(s_{T+1} = 0 | s_{T} = 1)$. In simulations, the attention state of a pedestrian at the next time step is determined by sampling from a single-trial binomial distribution: 
\begin{equation} \label{eq:bernoulli}
     s_{T+1} \sim \mathcal{B}(n=1, p=p(s_{T+1} = 1 | s_{T}))
\end{equation}

\subsubsection{The probability of attention state transition} \label{sec: inputs}

\begin{figure}[ht]
     \centering
     \begin{subfigure}[b]{0.22\textwidth}
         \includegraphics[width=\textwidth]{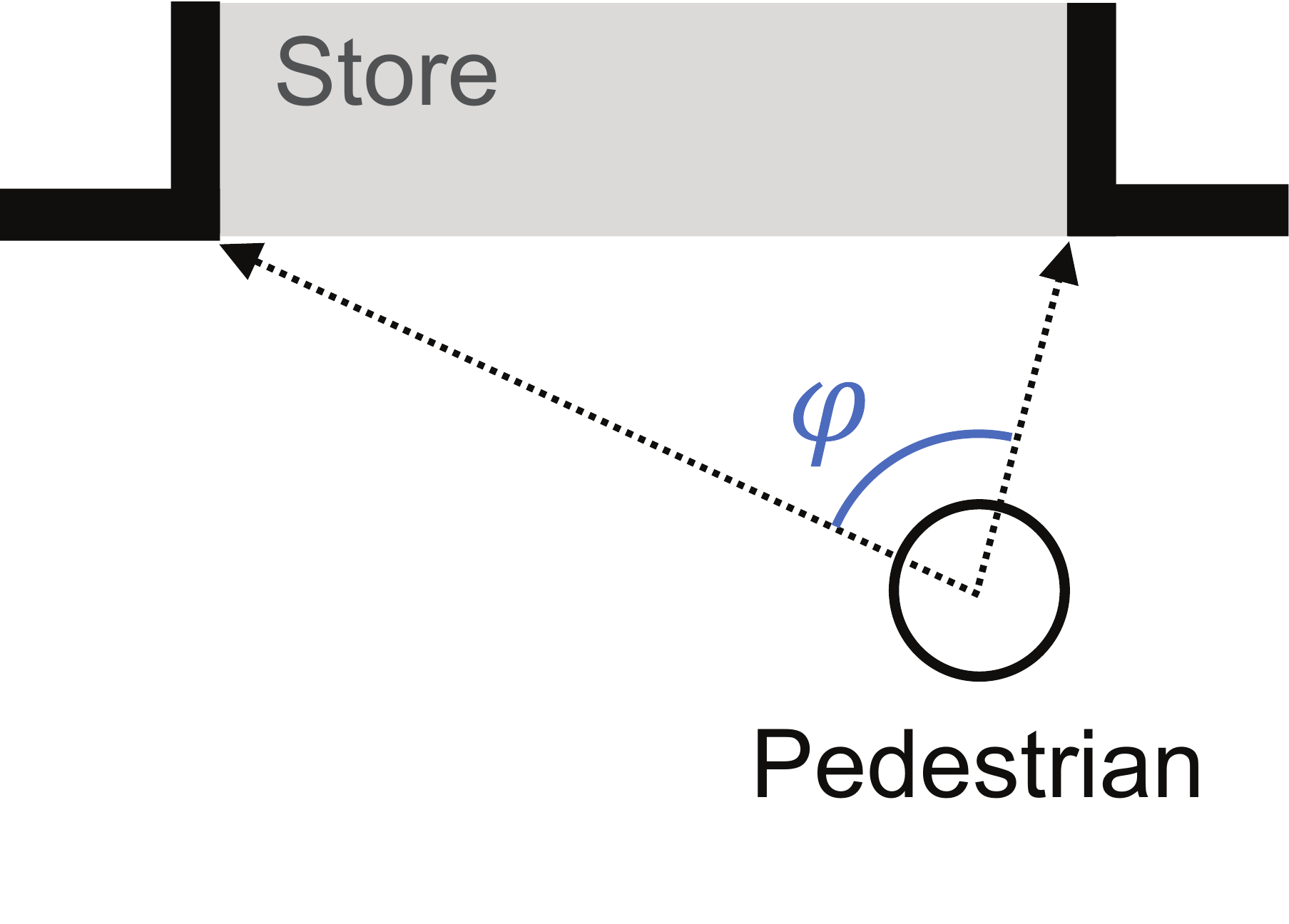}
         \caption{\cite{xie_signage_2007}}
     \end{subfigure}
     \begin{subfigure}[b]{0.22\textwidth}
         \includegraphics[width=\textwidth]{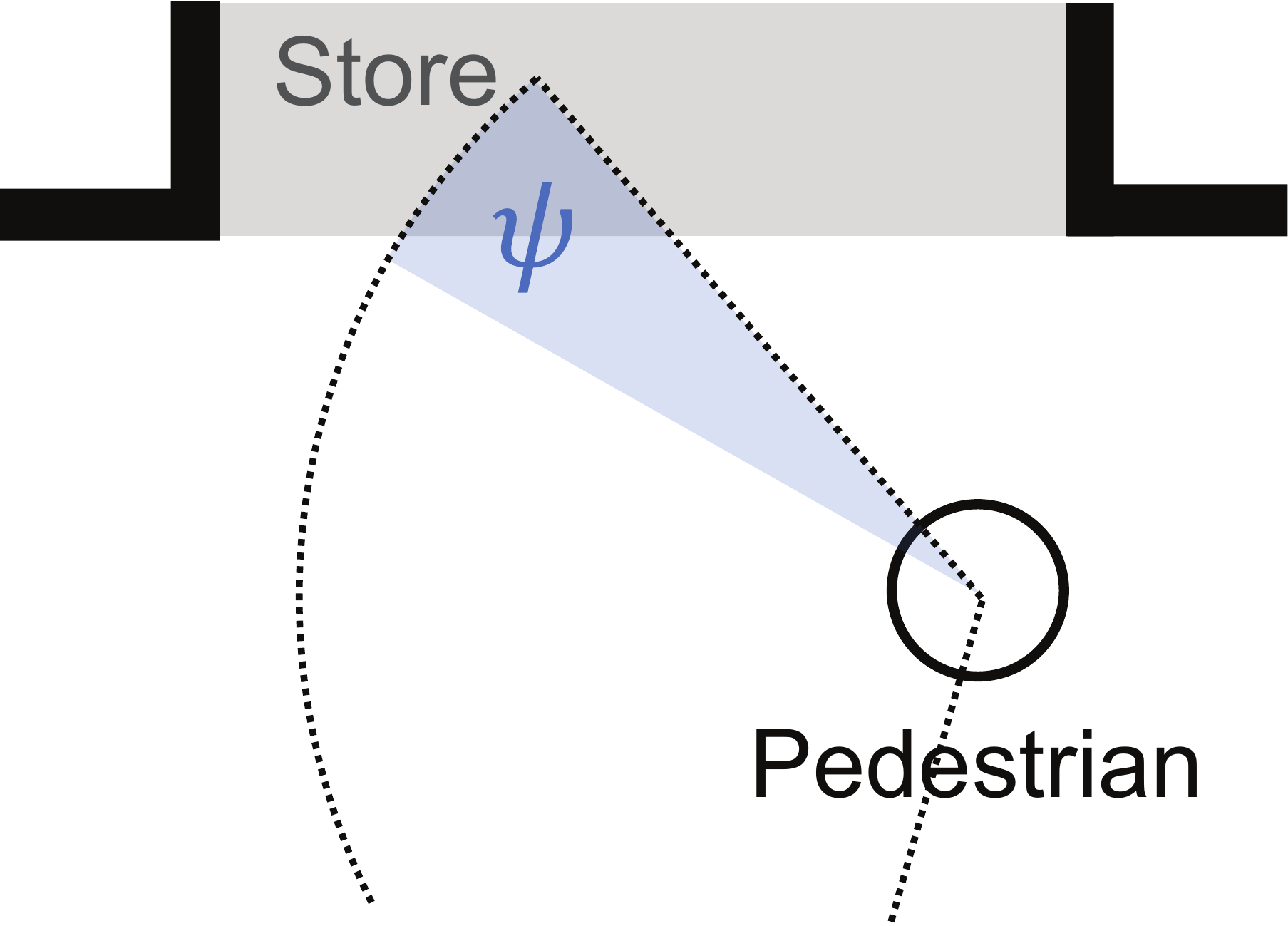}
         \caption{\cite{wang_microscopic_2014}}
     \end{subfigure}
     \begin{subfigure}[b]{0.22\textwidth}
         \includegraphics[width=\textwidth]{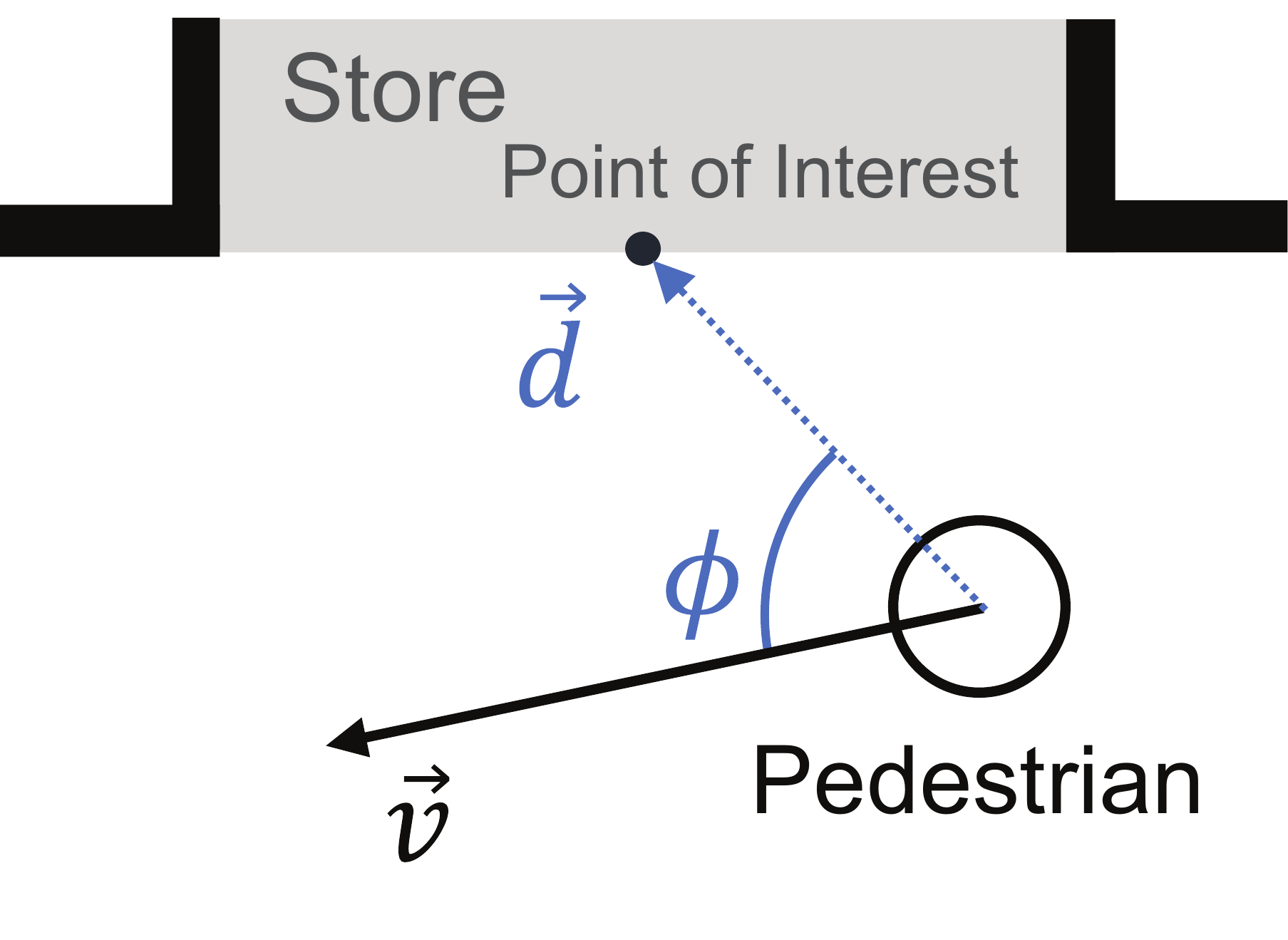}
         \caption{\cite{zhou_modeling_2022}}
     \end{subfigure}
     \begin{subfigure}[b]{0.22\textwidth}
         \includegraphics[width=\textwidth]{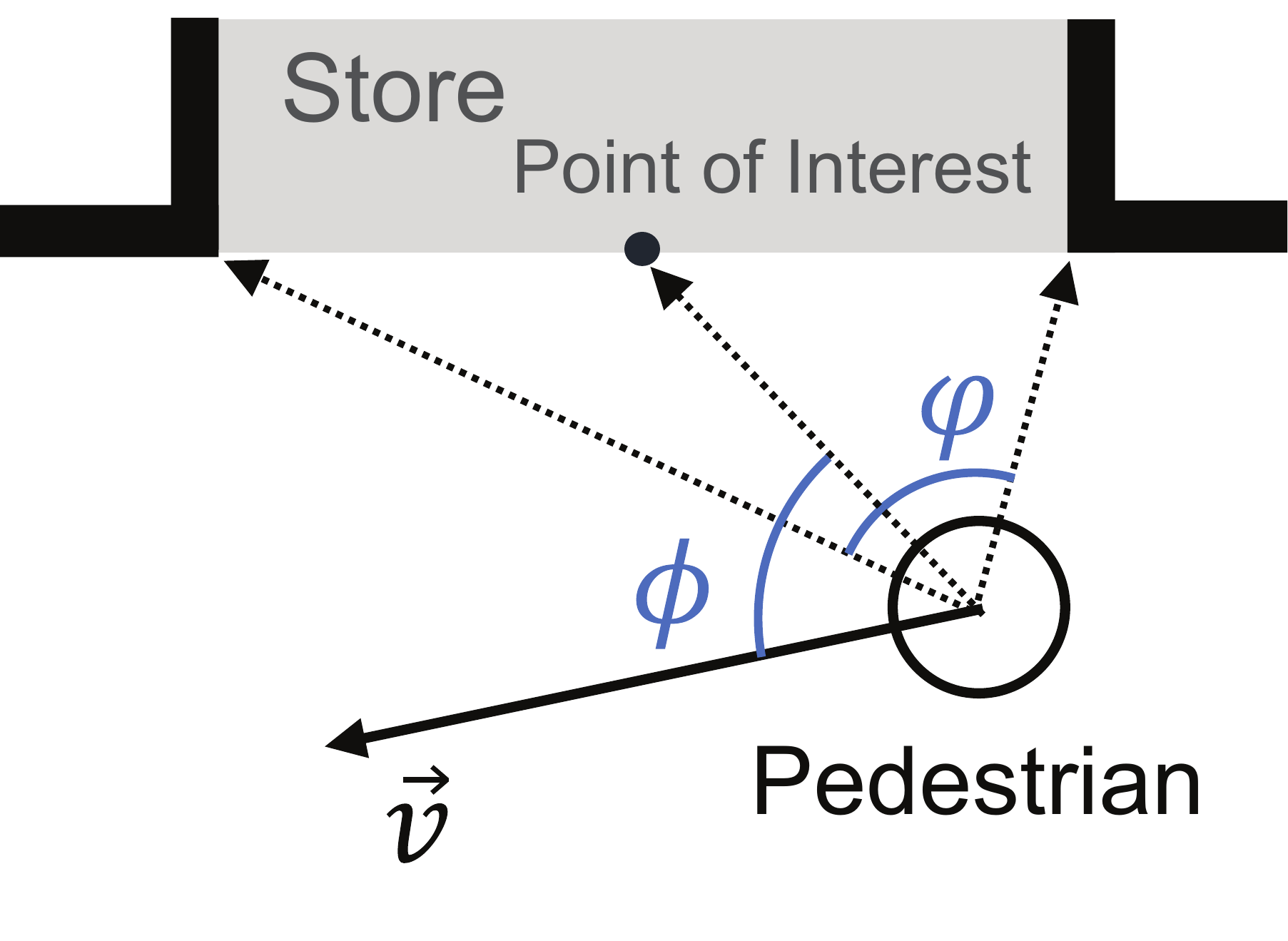}
         \caption{Our Model}
         \label{fig:our variable}
     \end{subfigure}

    \caption{\textbf{Input variables in prior and our work}. Generally, our model combines the observation angle $\phi$ in \cite{zhou_modeling_2022}. with the angular separation $\varphi$ in \cite{xie_signage_2007}, resulting in a representation that can deal with varying environmental object sizes and orientations.}
    \label{fig:input}
\end{figure}

To implement the relations between visual coverage information and visual attention, we convert the position $\vec{r}_T$ into two input variables below. Then, the probabilities of visual attention transitions are defined as functions of them. 

\begin{description}
\item[Angular Separation $\varphi$] Similar to the definition in \cite{xie_signage_2007}, the angular separation describes the size of coverage that the environmental object applies to the visual field of a pedestrian. From a top-down view of the walking environment, If two tangent lines are drawn from the pedestrian position to environmental object boundaries, the angle between the two is the angular separation (Fig. \ref{fig:input}).
\item[Observation Angle $\phi$] Similar to the definition in \cite{zhou_modeling_2022}, it is the angle between the walking direction and the direction pointing from the pedestrian to the environmental object. 
In Zhou's work, a point of interest represented an individual object (e.g., a sign). In contrast, our use case is an entire store. To reduce the large footprint of a store to a representative point, we select the mid-point of the store's entryway as the point-of-interest(see Fig. \ref{fig:our variable}). We use such a representation in both our model and our implementation of \cite{zhou_modeling_2022} in \S \ref{sec:evaluation-protocols} .
\end{description}

To keep the model in its possible simplest form but include the possible interaction effects between inputs, we choose polynomial logistic regressions to predict the binary output (attention state). To avoid over-fitting, we limit the highest order to 2 and conduct step-wise backward feature exclusion at the final step of the modelling. That is, starting from modelling visual attention with all variables and their high-order combinations, we remove terms one by one until the prediction performance begins to drop. So we have the probability of attention initiation (written in a form prior to step-wise backward feature exclusion):
\begin{equation} \label{eq:our_visual}
        p_{\theta}(s_T = 1 | s_{T-1} = 0) = \sigma(\theta_0 \varphi + \theta_1 \phi + \theta_2 \varphi ^ 2 + \theta_3 \phi ^ 2 + \theta_4 \varphi \phi + \theta_5)
\end{equation} 
and the probability of attention termination:
\begin{equation} \label{eq:our_visual2}
        p_{\theta}(s_T = 0 | s_{T-1} = 1) = 1 - p_{\theta}(s_T = 1 | s_{T-1} = 1) = 1 - \sigma(\theta_6 \varphi + \theta_7 \phi + \theta_8 \varphi ^ 2 + \theta_9 \phi ^ 2 + \theta_{10} \varphi \phi + \theta_{11})
\end{equation} 
where $\sigma(\cdot)$ denotes the sigmoid function. And all $\theta$s ranging from $\theta_0$ to $\theta_{11}$ are parameters to be fit. By plugging Eq. \ref{eq:our_visual} and Eq.\ref{eq:our_visual2} into Eq. \ref{eq:bernoulli}, the attention state is simulated as:
\begin{equation} \label{eq: visual}
s_{T+1} \sim \mathcal{B}(1, p) \text{ , where } p =
\begin{cases*}
    \sigma(\theta_0 \varphi + \theta_1 \phi + \theta_2 \varphi ^ 2 + \theta_3 \phi ^ 2 + \theta_4 \varphi \phi + \theta_5), & if $s_{T} = 0$,\\
    \sigma(\theta_6 \varphi + \theta_7 \phi + \theta_8 \varphi ^ 2 + \theta_9 \phi ^ 2 + \theta_{10} \varphi \phi + \theta_{11}), & if $s_{T} = 1$.
\end{cases*}
\end{equation}

Finally, to speed up the simulation and align with prior observation, we cut $s_{T+1}$ down to zero when the angular separation $\varphi$ is less than a small threshold. It means pedestrians will never draw attention to an object when it occupies a too-small area of their vision fields. The threshold is chosen as $0.29$ (in radian) from the empirical study in \cite{xie_signage_2007}. It should be noted that such a cut-off is only applied in simulation. And it imposes little impact on model fitting since the $\varphi$ in most datapoints are larger than the threshold.

\subsection{Locomotion Module} \label{locomotion}

This section describes the locomotion of attention-based movements. In the first two subsections, we analyze our empirical data with regard to direction and speed changes. In the third subsection, we demonstrate how the empirical findings are integrated into the Social Force Model. Finally, we show the parameters, specifications, and boundary conditions in the simulation.

\subsubsection{Weak direction changes} \label{sec: direction}

\begin{figure}[ht]
     \centering
     \begin{subfigure}{0.32\textwidth}
         \includegraphics[width=\textwidth]{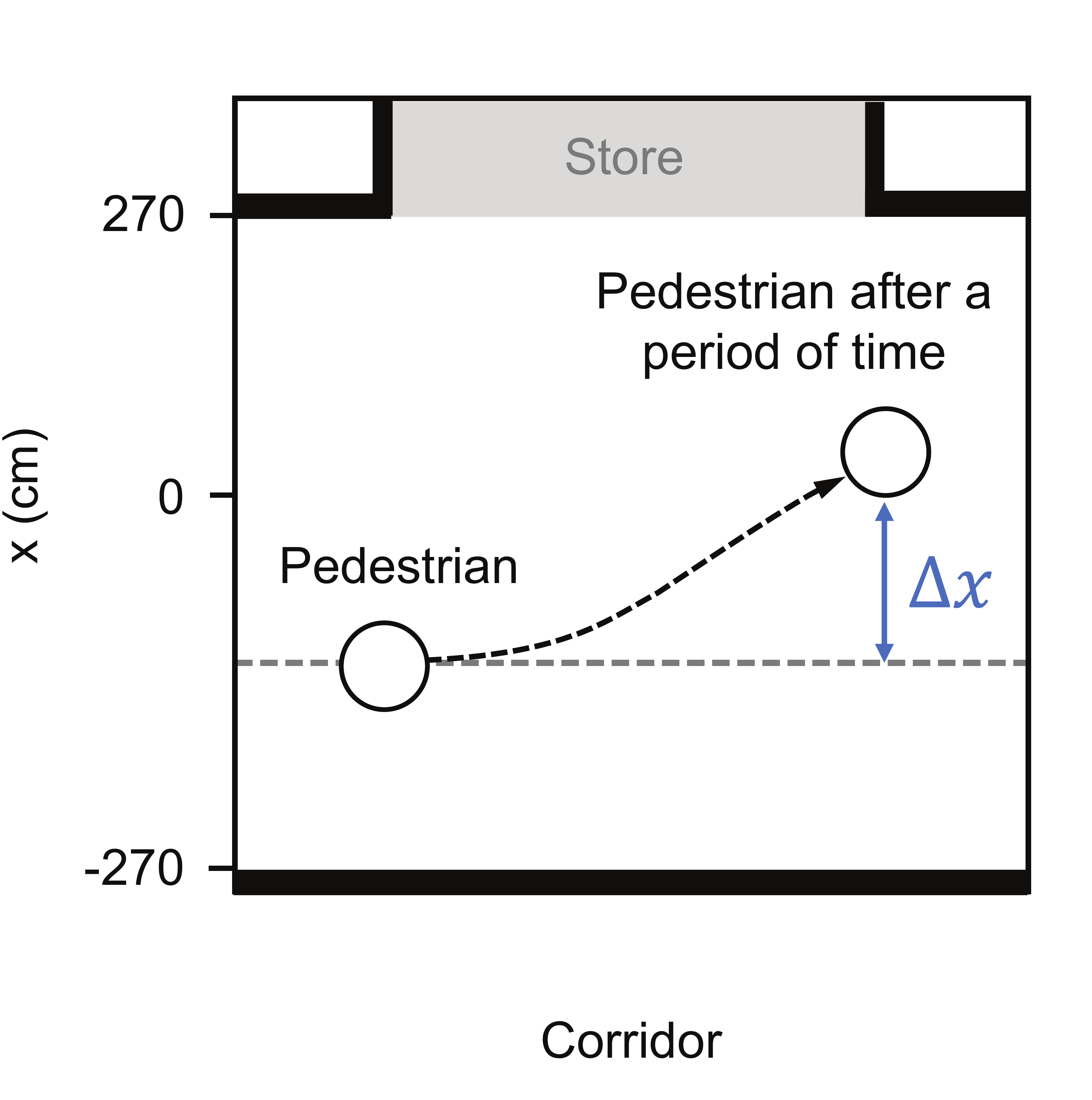}
         \caption{}
         \label{fig: delta x}
     \end{subfigure}
     \hfill
     \begin{subfigure}{0.32\textwidth}
         \includegraphics[width=\textwidth]{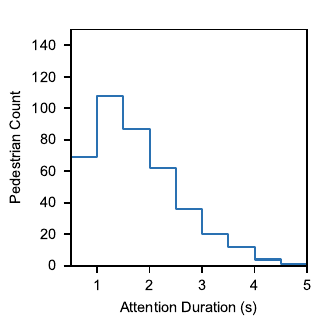}
         \caption{}
         \label{fig: count}
     \end{subfigure}
     \hfill
     \begin{subfigure}{0.32\textwidth}
         \includegraphics[width=\textwidth]{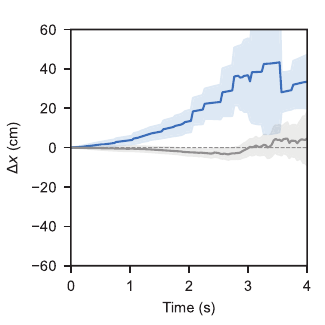}
         \caption{}
         \label{fig: lateral}
     \end{subfigure}

    \caption{\textbf{Walking direction change (lateral positional deviation) for pedestrians with visual attention}.
    (\subref{fig: delta x}) The illustration of walking direction change (lateral positional deviation) $\Delta x$.
    (\subref{fig: count}) The distribution of visual attention duration, shown in a histogram. Pedestrians without visual attention are not shown here.
    (\subref{fig: lateral}) The average deviation of lateral position after the initiation of attention. The grey solid line represents the deviation of pedestrians without visual attention after entering the camera.
    }
    \label{fig:loco}
\end{figure}

Given the pedestrians with attention movement behaviors will finally continue their original routes rather than walking into the store, we prefer measuring the accumulated impact of direction change instead of the difference between the first and the last time step. Therefore, we define walking direction change as the lateral positional deviation $\Delta x$ of a pedestrian after the initiation of visual attention. Fig. \ref{fig: delta x} depicts the definition.

We firstly explore how $\Delta x$ changes over time. Here we study all pedestrians who initiate their visual attention inside our camera view with a duration $\geq$ 1.5s. We set this threshold to lower the impacts of mislabelling attention states. As shown in Fig.~\ref{fig: count}, after 2.5s of the attention initiation, most pedestrians have terminated their attention. For remaining pedestrians, their average $\Delta x$ is significantly more than zero, but the value is only around 30cm (see Fig. \ref{fig: lateral}). Therefore, for the majority of pedestrians with visual attention, their lateral position deviation is small.

Secondly, we explore whether $\Delta x$ is affected by store proximity. We divide all pedestrians into $5$ groups by their lateral positions when attention is initiated. Then we compare how average $x$ changes across the groups. We observe no significant $\Delta x$ differences between groups after 1.5s of the attention initiation (F-test statistics=$0.32$ with the degrees of freedom (4, 112), p-value=$0.866$). And we also do not observe significant differences after 2.5s of the attention initiation (F-test statistics=$1.24$ with the degrees of freedom (3, 25), p-value=$0.317$). We are unable to compare groups with longer attention duration due to lack of data. While the insignificant results do not suggest store proximity applies no impact on $\Delta x$ in real world, they indicate the impact is small when attention duration is less than 2.5s.

Such an observation demonstrates that direction changes are very weak in attention-based movement behaviors. However, prior works \citep{kwak_collective_2013, zhou_modeling_2022} use additional SFM forces to describe similar behaviors. Additional forces are usually applied when pedestrians are assumed to steer towards environmental objects, and the steering effects increase (or decrease) with proximity. Since it is not aligned with our observation, we conclude that attention-based movements can not be well explained or simulated by additional SFM forces in prior works.

\subsubsection{Capped walking speed}
\label{sec: angular}

As direction changes do not play a key role in attention-based movement behaviors, speed changes are studied instead. Based on the idea in \cite{warren_behavioral_2008}, the angular speed with regard to the store display is studied. 

\begin{figure}[ht]
\centering
    \begin{subfigure}[t]{0.32\textwidth}
         \includegraphics[width=\textwidth]{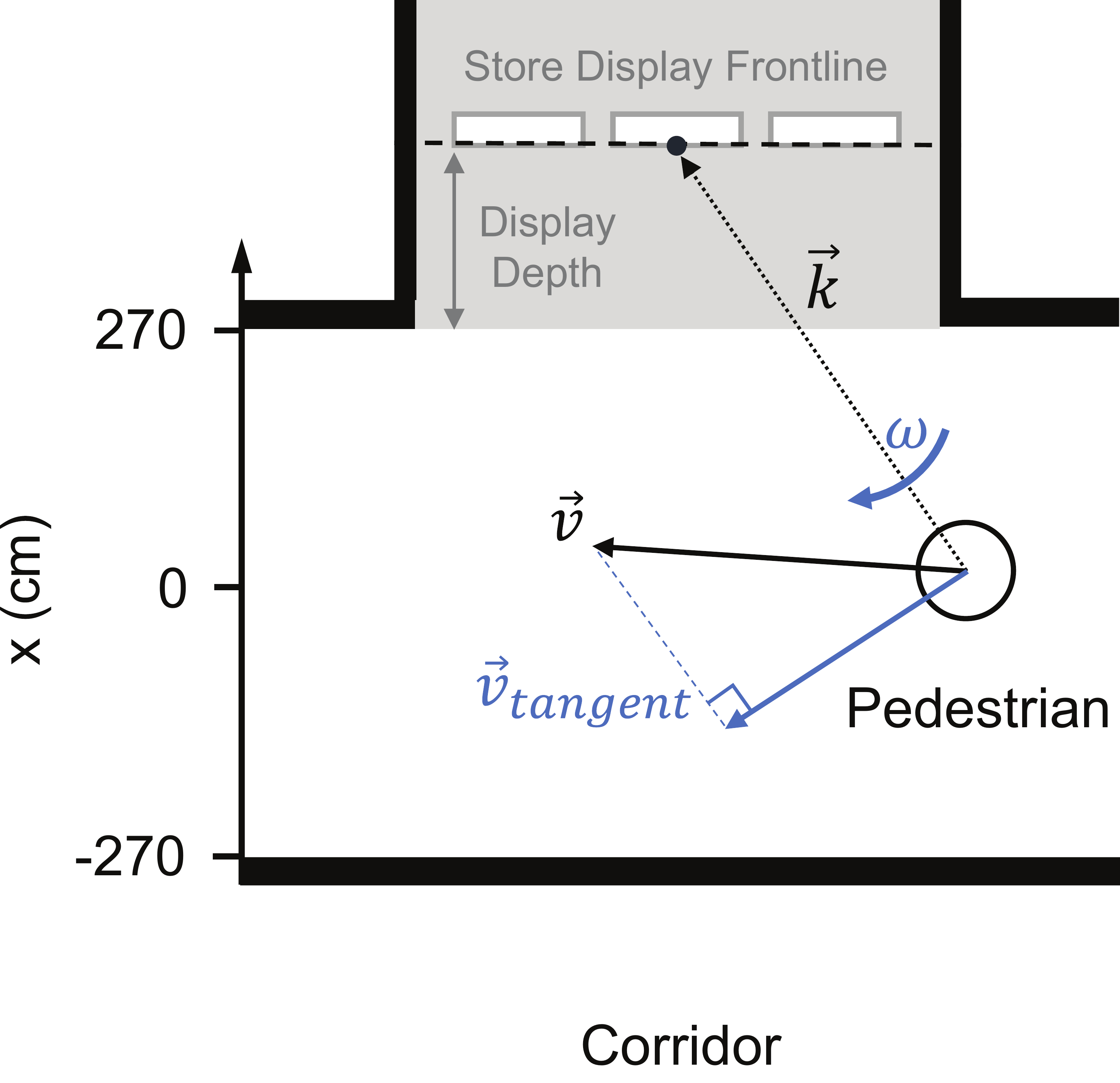}
         \caption{}
         \label{fig:angular}
     \end{subfigure}
     \begin{subfigure}[t]{0.32\textwidth}
         \includegraphics[width=\textwidth]{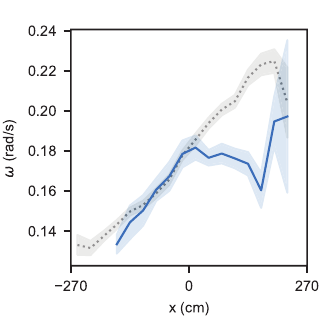}
         \caption{}
         \label{fig: omega}
     \end{subfigure}
     \begin{subfigure}[t]{0.32\textwidth}
         \includegraphics[width=\textwidth]{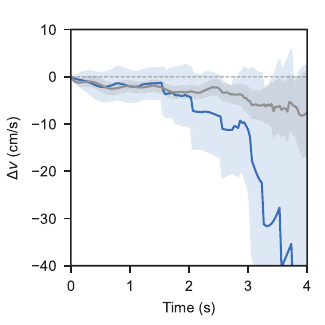}
         \caption{}
         \label{fig: v}
     \end{subfigure}
     
\caption{ \textbf{Angular speed $\omega$ to store display}. (\subref{fig:angular}) The illustration of angular speed $\omega$ and tangential velocity $\vec{v}_{tangent}$. (\subref{fig: omega}) How average angular speed $\omega$ varies with lateral position ($x$ coordinate). $x=0$ represents the middle of the corridor section. $x=270$ represents the wall boundary of the corridor on the store side and $x=-270$ represents the opposite boundary. The grey dotted line represents the angular speed of pedestrians without visual attention to the store. And the shaded areas represent 95\% CIs of the metrics.  (\subref{fig: v}) The average walking speed change after the initiation of attention. The grey solid line represents the speed change of pedestrians without visual attention after entering the camera. The shaded areas represent 95\% CIs of the metrics.}

\end{figure}

Here the store display refers to goods that are visible to pedestrians outside the store. Such items would be clothes in the window display or food on shelves. While the layouts of store displays vary, the convenience store in our study adopts a simple strategy, where items are stacked on rows of shelves that form a frontline to pedestrians outside (see Fig. \ref{fig:angular}). In this way, we represent the store display by a frontline. This representation can be extended to other displays such as window displays, where the display depth (the distance from the frontline to the store entrance) is small. Next, the position of the store display is further defined as the midpoint of the store display frontline (see Fig. \ref{fig:angular}). It should be noted that this midpoint for store display is different from that for the store in Fig.\ref{fig:our variable}.

Then the angular speed to the store display is calculated in this way: firstly we construct a vector $\vec{k}$ pointing from the pedestrian to the midpoint of the store display. Then a tangential velocity $\vec{v}_{tangent}$ is derived from projecting walking velocity $\vec{v}$ onto the direction that is perpendicular to $\vec{k}$. Finally, the angular speed $\omega$ is the $\vec{v}_{tangent}$ divided by the length of $\vec{k}$.

To explore whether the angular speed is related to the proximity to the store, the average angular speeds stratified by pedestrian lateral positions are compared. The average angular speed is calculated in this way: for a given lateral position stratum, we query all datapoints falling into the stratum. Then the datapoints are grouped by pedestrian IDs to get the average angular speeds for each pedestrian. Finally, the average angular speeds are averaged across all pedestrians.

The result shows that pedestrians with visual attention to the store differ from those without attention in their angular speed. As shown in Fig. \ref{fig: omega}, while the average angular speed for pedestrians without attention increases with the proximity to the store entrance (grey dotted line), the angular speed for pedestrians with visual attention (blue line) is capped by a threshold value (its mean $\meanidealangular$ and standard deviation $\varianceidealangular$ are included in Tab. \ref{subtab:locomotion-regulation}). As a result, pedestrians whose attention is drawn to the store will walk more slowly, with the speed decreasing more as their distance to the store reduces.

We also explore how walking speed varies over time after the attention initiation. However, we do not observe significant speed differences for pedestrians who initiate visual attention inside camera view(see Fig. \ref{fig: v}) due to lack of data. Although we cannot reach a thorough conclusion on speed change over time, we can still use the current observation on the angular speed to build a locomotion model that approximates the slowing dynamics.

\subsubsection{Social Force Model Integration} \label{sec: sfm integration}

The general idea of the locomotion module is: if pedestrians draw their visual attention to the store, they will adjust their desired speed to ensure the angular speed towards the store display will not exceed a certain value. We firstly explain how the desired speed regulates locomotion using Social Force Model. Then we demonstrate how to design the dynamic desired speed function. 

In the Social Force Model (using the notation in equation (2) and (3) in \cite{johansson_specification}), for a pedestrian $\alpha$, its velocity $\vec{v}_{\alpha}$ is calculated by:
\begin{equation} 
        \frac{d \vec{v}_{\alpha}(t)}{dt} = \frac{1}{\tau_{\alpha}} (v^0_{\alpha}\vec{e}_{\alpha} - \vec{v}_{\alpha}) + \sum_{\beta (\alpha)} \vec{f}_{\alpha \beta}(t) + \sum_{i} \vec{f}_{\alpha i}(t) + \vec{\xi}_{\alpha}(t)
        \label{eq: sfm}
\end{equation}
where $f_{\alpha \beta}(t)$ and $\vec{f}_{\alpha i}$ denote the repulsive effects from other pedestrians and the environment, respectively. $\vec{\xi}_{\alpha}(t)$ is a random noise attribute. The first term (the goal force), describes the tendency to adjust the current velocity $\vec{v}_{\alpha}$ to a desired one $v^0_{\alpha}\vec{e}_{\alpha}$. It should be noted that $v^0_{\alpha}$ is the desired speed in scalar and $\vec{e}_{\alpha}$ is the desired direction (vector) pointing to the pedestrian's goal.

In \cite{helbing1995social} and \cite{johansson_specification}, while the desired speed $v^0_{\alpha}$ varies among pedestrians, it is a constant for a single pedestrian over time. In our paper, we turn it into a dynamic one. We update the desired speed at every time step by scaling a neutral speed with the ratio between an ideal angular speed and the current angular speed. And the ratio is capped by $1$. As a result, if the current angular speed is greater than the ideal angular speed, the ratio will be less than $1$. Then a smaller desired speed will guide a pedestrian to slow down. If the current angular speed is smaller than the ideal, the ratio will still be $1$ so no influence is applied. The calculation steps are detailed below.

We first calculate the current angular speed by dividing the magnitude of tangential speed $\vec{v}_{tangent}$ with the length of $\vec{k}$ (illustrated in Fig. \ref{fig:angular}):
\begin{equation} 
        \omega_{current} = \frac{\|\vec{v}_{tangent}\|}{\|\vec{k}\|} = \left( 1-\frac{\vec{v} \cdot \vec{k}}{\|\vec{v}\| \cdot \|\vec{k}\|} \right) \cdot \frac{\|\vec{v}\|}{\|\vec{k}\|}
        \label{eq:tangential_v}
\end{equation}

Then we calculate the ratio $\zeta$ between the ideal angular speed $\omega_{ideal} \sim \mathcal{N}(\meanidealangular, \varianceidealangular^2)$ (detailed in \S \ref{sec: angular}) and the current angular speed. $\omega_{ideal}$ is sampled once when a pedestrian is initiated in the simulation. The ratio is capped by $1$.
\begin{equation} 
        \zeta = \min(\omega_{ideal} / \omega_{current}, 1)
\end{equation}

Next, the desired speed is scaled by the ratio. Besides, such a scaling only happens when the visual attention has been initiated (i.e. $s_{T+1} = 1$):
\begin{equation} 
    v^0_{\alpha} = v_{neutral} \cdot \left[\zeta \cdot s_{T+1} + (1-s_{T+1})\right]
    \label{eq: v_desire}
\end{equation}
where $v_{neutral}$ represents the desired speed without visual attention. It is fixed for a single pedestrian over time. But it varies among pedestrians, which is elaborated in \S \ref{sec:boundary}.
In this way, the desired speed controlled by attention states acts as a dynamic term in the goal force component of the Social Force Model.

\subsubsection{Simulation configurations} \label{sec:boundary}

To demonstrate that the success of our model relies on the design of visual attention module and its locomotion regulation (rather than the overfitted parameters in SFM and boundary condition), we configure SFM and the boundary condition only using pedestrians that don’t show visual attention. In other words, we split all pedestrians in the empirical dataset into 3 disjoint groups: (1) pedestrians that never initialize visual attention or walk into the store, i.e. commuters. (2) pedestrians that initialize visual attention, that is, having attention-based movement behaviors. (3) pedestrians that enter or exit the store, i.e. consumers. We only use the first group for SFM and boundary condition (except flow rates) configuration. And we use first two groups to fit visual attention module and the flow rates. Then, we run two simulations with visual attention module activated and deactivated respectively. We expect that the simulation without visual attention module should produce results that resemble the behaviors of commuters (the first group). And the one with visual attention should be similar to the union of first two groups. The third group is not in our scope, as we clarified in \S \ref{sec:intro}. The results are shown in Fig. \ref{fig:v_emp_and_sim_wo_attn} and \ref{fig:v_emp_and_sim_w_attn}. And the configuration details are provided as below:

\paragraph{Social Force Model} We do not fit the Social Force Model (locomotion module) ourselves because there are few neighbouring pedestrians for a large proportion of our trajectories, which lead to insufficient data for a traditional model fitting. Instead, we adapt existing models and specifications in prior work.
Given that prior models were fitted using dataset with different pedestrian flow densities, we assign different parameters and specifications for different flow density scenarios. 
In this paper, we consider the cases where the pedestrian Level-of-Service is no more than Level C ($1.4 m^2$ per person, equivalent to a flow rate at around $0.5 people/(m \cdot s)$ in all directions in our paper) as low-density scenarios, otherwise as high-density.

For low-density simulations that appear in the model validation \S \ref{sec:model-validation} and the first case study \S \ref{sec:case1}, we adapt the parameters and specifications in \cite{zanlungo_microscopic_2012}, since it is fitted for such scenarios and is capable of replicating the speed distribution along the walking facility section \citep{zanlungo_microscopic_2012}. Based on its original parameter values, we change two of them ($A$ and $\tau^{-1}$, see Tab. \ref{subtab:sfm}) to let the walking speed distribution to be aligned with the empirical data (see Fig. \ref{fig:v_emp_and_sim_wo_attn}). For high-density simulations that appear in the second case study \S \ref{sec:case2}, we adopt \cite{johansson_specification}, which does better in maintaining pedestrian flow stability in high-density scenes.

\paragraph{Boundary condition} We set the boundary conditions of pedestrian flows to resemble the original dataset in ~\cite{Li2022}. The boundary conditions are set by three key attributes: pedestrian flow rate, neutral speed, and density distribution. They are analysed and modeled as below:

\begin{figure}[ht]
     \centering
     \begin{subfigure}[b]{0.23\textwidth}
         \includegraphics[width=\textwidth]{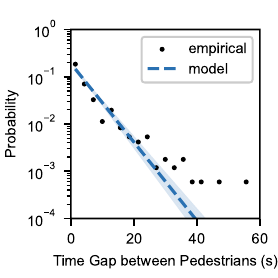}
         \caption{}
         \label{fig: bound_flow_rate}
     \end{subfigure}  
     \begin{subfigure}[b]{0.23\textwidth}
         \includegraphics[width=\textwidth]{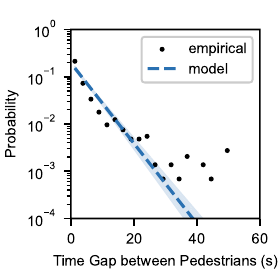}
         \caption{}
         \label{fig: bound_flow_rate2}
     \end{subfigure}    
     \begin{subfigure}[b]{0.23\textwidth}
         \includegraphics[width=\textwidth]{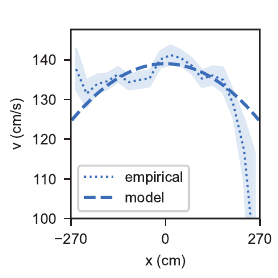}
         \caption{}
         \label{fig: bound_speed}
     \end{subfigure}  
     \begin{subfigure}[b]{0.23\textwidth}
         \includegraphics[width=\textwidth]{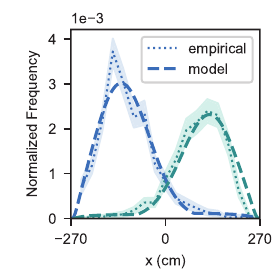}
         \caption{}
         \label{fig: bound_density}
     \end{subfigure}

    \caption{\textbf{Boundary conditions}: (\subref{fig: bound_flow_rate}) and (\subref{fig: bound_flow_rate2}) shows the probability distributions of the time gaps between pedestrians in two directions. The dots are empirical measurements. The dashed lines are exponential fitting results. The y-axes are shown in log scales. (\subref{fig: bound_speed}) shows how the average walking speed varies with lateral coordinates for pedestrians without visual attention, suggesting that pedestrians close to walls tend to walk more slowly. (\subref{fig: bound_density}) shows the density distributions of pedestrians as a function of lateral coordinates. The two distributions in two colors represent the pedestrian flows in two directions. The shaded areas represent the 95\% CIs for the metrics.}
    \label{fig:boundary}
\end{figure}

The pedestrian flow rate, referring to the number of pedestrians moving through a designated area over a period of time, is represented by time gaps between two pedestrians crossing the same vertical plane. After removing the gaps over 60 seconds, the time gaps follow an exponential distribution where the mean values are detailed in Tab. \ref{subtab:boundary} (distributions are illustrated in Fig. \ref{fig: bound_flow_rate} and \ref{fig: bound_flow_rate2}).

For pedestrians who never have attention drawn to the store, their walking speed distribution is used as the mean of the neutral walking speed $\avgneutralspd$ for all agents. Fig.~\ref{fig: bound_speed} shows that pedestrians tend to walk faster at the center of the corridor. Such a trend is approximated by a second-degree polynomial function, where the mean of the neutral speed is affected by the initial lateral position: $\avgneutralspd(x) = a_vx^2 + b_v$,  with $x$ in centimeters and $\avgneutralspd$ in centimeters per second. When agents are initialized, their desired walking speeds without visual attention $v_{neutral}$ are sampled from $\mathcal{N}(\avgneutralspd(x), \sigma_v^2)$, where the standard deviation $\sigma_v$ is also calculated from the observation data.

Finally, the density function of pedestrians' initial lateral position is modeled the same as a prior work \citep{zanlungo_experimental_2014}. In \cite{zanlungo_experimental_2014}, the density function is proportional to a Boltzmann factor: $p(x) \propto e^{U(x)}$. And the energy function $U(x)$ is designed to resemble a Gaussian at its maximum, drop to zeros at wall boundaries, and maintain a constant value far enough from the walls and its peak:
\begin{equation*} 
        U(x) = \frac{a_{\rho}}{x} + \frac{a_{\rho}}{L_c-x} + (\frac{\delta}{b_{\rho}L_c})^2
\end{equation*}
\begin{equation*} 
\begin{array}{rcl}
        \delta &=& \begin{cases}
            x-c_{\rho}L_c& \text{if $|x - c_{\rho}L_c| \leq d_{\rho}L_c$} \\
            d_{\rho}L_c& \text{if $|x - c_{\rho}L_c| \geq d_{\rho}L_c$}
        \end{cases}
    \end{array}
\end{equation*}

\noindent where $L_c=540cm$ is the width of the corridor. $a_{\rho}$ is fixed to 30cm, as indicated in \cite{zanlungo_experimental_2014}. The remaining parameters are calibrated using grid search with our dataset. The definition and the calibration results of these parameters are summarized in Table \ref{subtab:boundary}. The fitted distribution and empirical measurements are visualized in Fig.~\ref{fig: bound_density}.

\subsection{Evaluation Protocols} \label{sec:evaluation-protocols}

Our model is evaluated in two steps: firstly we compare the attention prediction results to ground truth, where we measure the performance of the visual attention module solely. Then we compare the simulation results to ground truth to know the performance of the whole model.

\subsubsection{Attention Metrics}
We evaluate the visual attention module by three aspects. First, to prove that we chose powerful variables as model inputs, our representations are compared to other possible combinations of variables in previous literature. Second, our model is compared to visual-attention models in selected previous work~\citep{wang_microscopic_2014, zhou_modeling_2022, kremer_watch_2020} to prove its performance in attention transition events (initiation and termination). Finally, as prediction errors for individual time steps may accumulate over time, good prediction performance on transition events may not necessarily mean good performance on sequential predictions. Therefore, to evaluate such accumulated errors, we compare the probability distributions of attention duration among all models.

\begin{table}[h]
\caption{Variables for Representation Evaluation. It should be noted that some variables ($\psi$, $d$, $n_g$, $d_{ir}$, $\rho$, $t_g$, $a_{max}$) are neither parts of our visual attention module nor locomotion module. They are only used for representation evaluation that is shown in \ref{subsec:resultAttention}. }\label{tab:inputs}
\begin{tabular*}{\tblwidth}{@{}LCL@{}}
\toprule
Variable & Notion & Definition\\
\midrule
\textit{Intrinsic Variable}&&\\
 Angular separation & $\varphi$ & \cite{xie_signage_2007}, also see \S \ref{sec: inputs}\\ 
 Observation angle & $\phi$ & \cite{zhou_modeling_2022}, also see \S \ref{sec: inputs} \\  
 Visibility & $\psi$ & \cite{wang_microscopic_2014}\\ 
 Observation distance & $d$ & \cite{zhou_modeling_2022}\\ 
 Walking speed & $v$ \\
 Angular speed & $\omega$ & \begin{tabular}[t]{@{}l@{}}angular speed with regard to environmental objects, \\ see \S \ref{sec: angular} for details\end{tabular}  \\
 Cumulative gazing time & $t_g$ & the cumulative time of gazing for a pedestrian\\
Walking direction & $d_{ir}$ & \begin{tabular}[t]{@{}l@{}}a binary variable (0 or 1) for walking directions\\ in bi-directional corridors\end{tabular}\\
\midrule
\textit{Extrinsic Variable}&&\\
 Gazing pedestrian number & $n_g$ & \begin{tabular}[t]{@{}l@{}}
              the number of gazing surrounding pedestrians\\ whose observation angle 
              is less than 120 degrees\\ and the distance is less than 3m
          \end{tabular} \\
 Pedestrian density & $\rho$ & the X-T definition \cite{duives_analysis_2016} adapted from \cite{edie_discussion_1963}\\ 
Maximum surrounding acceleration& $a_{max}$ &\begin{tabular}[t]{@{}l@{}}
              the maximum acceleration of surrounding pedestrians.\\ The scope of "surrounding" is same as that in \\ gazing pedestrian number $n_g$
          \end{tabular} \\
\bottomrule
\end{tabular*}
\end{table}

\begin{description}
\item[Representation Evaluation] A list of independent variables (intrinsic and extrinsic) is derived from a literature review. They are defined and calculated as in previous studies (see Table.~\ref{tab:inputs}). To provide a general impression of variable impacts, we enumerate all uni-variate and bi-variate combinations for all listed variables. And we use those combinations to construct logistic regression models with the highest order up to $2$. Then we compare the prediction performances of all models.

\item[Evaluation on transition events] Since previous work is not constrained to logistic regression, our model is compared to the works of ~\cite{wang_microscopic_2014, zhou_modeling_2022, kremer_watch_2020} by fitting and validating with our dataset. All models are fit using gradient descent with an Adam optimizer. However, for \cite{wang_microscopic_2014} and \cite{kremer_watch_2020}, we also apply grid-search for comparison as their model structures are not designed for gradient descent and suffer exploding gradients in certain cases. Although we have applied smoothing terms to mitigate their numerical instabilities in gradient-descent, we believe their models may exhibit larger potentials if they are fit by grid-search.
\end{description}

Both evaluation tasks follow the same procedure: the datapoints in the original empirical dataset are firstly down-sampled at a sampling rate of 6Hz. After the standardization, the sampled data is split into a training set and a testing set at a ratio of 3:1. Next, models are fit to the training set. Evaluation metrics are derived by running prediction tasks on the testing set. All procedures above are repeated (100 times for representation evaluation and 30 times for the other) to get the distribution of the metrics. ROC Curves and AUC scores are provided to describe the true positive rate when the false positive rate is controlled. The results using this protocol are shown in \S~\ref{subsec:resultAttention}.

\begin{description}
\item[Evaluation on attention duration] As we assume a model with fewer accumulated errors produces an attention duration distribution that is closer to empirical observation, we calculate the difference between the attention duration distribution of empirical data and that of simulation for each model. The duration distribution for each model is derived by simulating all attention states of each pedestrian trajectory frame-by-frame where the previous attention state for each frame is the previous simulation result. On the contrary, in the previous two evaluation tasks, previous attention states were always the ground truth. The difference between two duration distributions is shown by the difference of their cumulative distribution functions (CDFs). Wasserstein distance, which sums up the absolute of these differences, was chosen as the quantitative metric. The results are shown in \S~\ref{subsec:resultAttention}.
\end{description}

\subsubsection{Locomotion Metrics} \label{sec:model-validation}

To demonstrate that our model can jointly predict visual attention and speed variations, which are proxies of retail potential and pedestrian efficiency, the proportion of long-attention pedestrians and the average speed are compared to empirical datasets. In addition, to demonstrate our model's ability to show the spatial variance of these metrics, the metrics are calculated as functions of lateral position. They are defined as below:

\begin{description}
\item[Proportion of Long-attention Pedestrians $P_{long}(X)$] Pedestrians who draw their attention to the store for no less than $2.5$s are defined as long-attention pedestrians. The duration threshold here is chosen based on the empirical observation in \cite{gidlof_looking_2017}, where the probability of buying starts to rapidly increase when pedestrians spend more than $2.5$s looking at a product. Then, given a lateral coordinate stratum $X \in [a,b)$, we count the number of pedestrians whose datapoints fallen into this stratum as $N_{all}$. Next, we count the number of long-attention pedestrians in this stratum as $N_{long}$. Finally we have $P_{long}(X) = N_{long} / N_{all}$.
\item[Walking Speed $v(X)$] For each pedestrian $p$ in the population $S$, for all its datapoints falling into the lateral coordinate stratum $X \in [a,b)$, average the walking speed as $v_{p, X}$. After we get all $v_{p, X}$ for every $p$, the final result $v(X)$ is the average of all $v_{p, X} > 0$.
\end{description}

\section{Results}
\label{sec:4}

In this section we firstly show the prediction performance of our visual attention module. Then we demonstrate that our entire framework can simulate the attention-based movement with regrad to both visual attention and locomotion.

\subsection{Visual Attention Modeling} \label{subsec:resultAttention}

\begin{figure}[ht]
     \centering
     \begin{subfigure}[b]{0.35\textwidth}
         \centering
         \includegraphics[width=\textwidth]{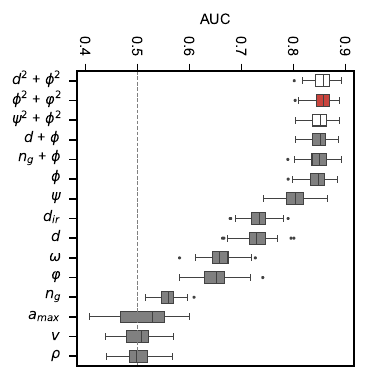}
         \caption{Attention Initiation}
     \end{subfigure}
     \begin{subfigure}[b]{0.35\textwidth}
         \centering
         \includegraphics[width=\textwidth]{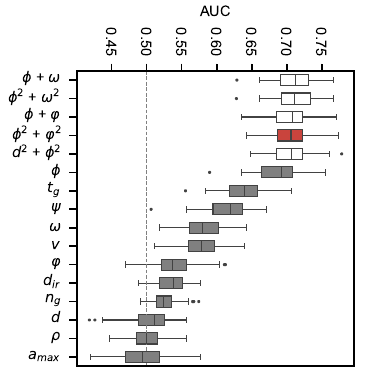}
         \caption{Attention Termination}
     \end{subfigure}
    \caption{\textbf{AUC scores of visual attention predictions with different representations}. Boxplots are drawn based on 100 trials with randomized samplings of our dataset. An AUC score equal to 1 indicates the model is always correct in the test dataset. And a score close to 0.5 indicates that the model performance is the same as random choices. The red boxes denote the performance of our representation. The grey boxes are representation performances that have significantly different mean values from ours. The x-axis labels are abbreviations of representations, where a single letter means a uni-variate linear logistic regression. And two letters connected by a plus (such as $\phi + \omega$) mean a bi-variate linear logistic regression. A squared letter means the logistic regression is polynomial up to 2 degrees. It should be noted that we only display the top five results and all uni-variate linear fitting results here.}
    \label{fig:auc}
\end{figure}

In the representation evaluation, our model representation was among the top candidates with regard to prediction performance, whereas other top candidates showed no significant difference from ours (see Fig.~\ref{fig:auc}).

Among all uni-variate logistic regressions, observation angle $\phi$ was the most influential variable in both attention state transition events. On the contrary, some variables (mostly extrinsic ones) imposed little impact on the dataset. For instance, the AUC scores of pedestrian density $\rho$  and maximum surrounding acceleration $a_{max}$ on both attention transition events were close to $0.5$ (see Fig.~\ref{fig:auc}). Also, gazing pedestrian number $n_g$ (defined in Table \ref{tab:inputs}) played little role in attention termination predictions. However, the results do not indicate the extrinsic variables are not influential in all real-world cases. Because our dataset did not capture pedestrian flows with extreme densities.

\begin{figure}[htbp]
     \centering
     \begin{subfigure}[b]{0.35\textwidth}
         \centering
         \includegraphics[width=\textwidth]{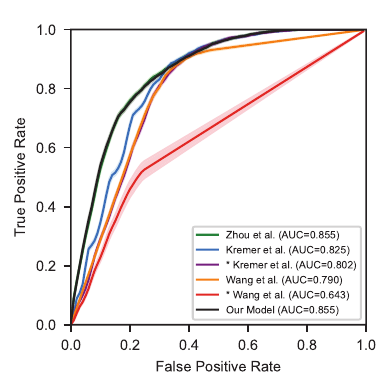}
         \caption{Attention Initiation}
         \label{fig:roc_start}
     \end{subfigure}
     \begin{subfigure}[b]{0.35\textwidth}
         \centering
         \includegraphics[width=\textwidth]{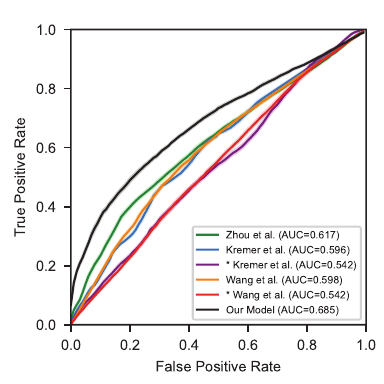}
         \caption{Attention Termination}
         \label{fig:roc_stop}
     \end{subfigure}
     \begin{subfigure}[b]{0.35\textwidth}
         \centering
         \includegraphics[width=\textwidth]{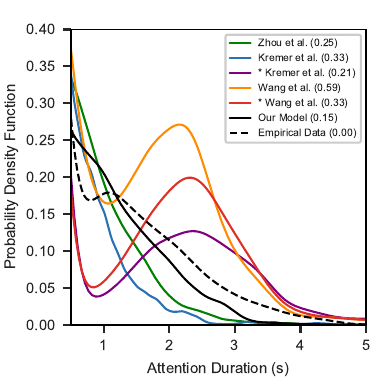}
         \caption{Attention Duration Distribution}
         \label{fig:pdf_attention}
     \end{subfigure}
     \begin{subfigure}[b]{0.35\textwidth}
         \centering
         \includegraphics[width=\textwidth]{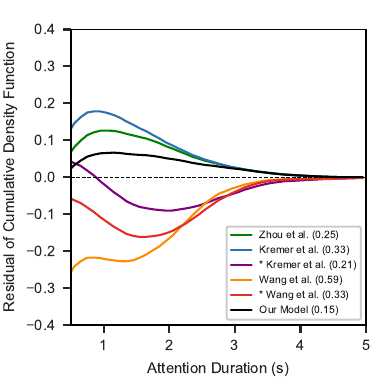}
         \caption{Residual of Duration Distribution}
         \label{fig:cdf_attention}
     \end{subfigure}
     
    \caption{\textbf{Compare our visual attention module with prior work}. (\subref{fig:roc_start}) and (\subref{fig:roc_stop}) are ROC curves with regard to attention initiation and termination predictions. When the curve gets closer to the top-left corner, the model achieves a higher true positive rate when the false positive rate is fixed. Shaded areas represent the standard errors of true positive rates. Our model is on par with \cite{zhou_modeling_2022} in attention initiation modelling, shown by their overlapping curves. The model names with asterisks (e.g., \texttt{*Wang et al.}) refer to the modified models fitted by gradient descent. (\subref{fig:pdf_attention}) compares the attention duration distributions of model results and empirical data by probability density functions. For visualization quality, the lines are smoothed by a Gaussian kernel. In the legend, the numbers in the parenthesis are the Wasserstein distances to the empirical distribution. (\subref{fig:cdf_attention}) shows the differences between the attention duration distribution of model simulation and that of empirical data, represented in the residuals of cumulative distribution function. When the line is closer to y=0, the simulation result is closer to empirical data. The legend is the same as the previous sub-figure, and lines are also smoothed by a Gaussian kernel. 
    }
    \label{fig:roc}
\end{figure}

Next, we compare our model with prior work in visual attention modelling. For attention initiation events (see Fig.~\ref{fig:roc_start}), our model was on par with~\cite{zhou_modeling_2022}, a reasonable result given our representation is a coordinate transform of \cite{zhou_modeling_2022} for the attention initiation part. For attention termination events (see Fig.~\ref{fig:roc_stop}), our model achieved $11\%$ higher AUC score than the best prior work. With regard to attention duration, our model achieved the closest distribution to empirical observation ($40\%$ less Wasserstein distance than the best prior work, see Fig.~\ref{fig:cdf_attention}). While \cite{zhou_modeling_2022} and \cite{kremer_watch_2020} tend to overestimate the proportion of pedestrian with short attention duration, ours excels in predicting long-attention pedestrians, which increases the overall performance. Although we introduce more parameters than some previous studies (\cite{wang_microscopic_2014, kremer_watch_2020}), it achieves prediction improvements without significant additional computational cost.

Finally, using the whole dataset, our final model fitting results are shown in Table \ref{subtab:visual-attention}. It should be noted that some variable combination terms (such as $\theta_8$ for $\varphi^2$ in attention termination predictions) were removed from Eq.\ref{eq: visual} by step-wise backward feature selections.

\subsection{Locomotion Modeling}
\begin{figure}[htbp]
     \centering
     \begin{subfigure}[t]{0.3\textwidth}
         \includegraphics[width=\textwidth]{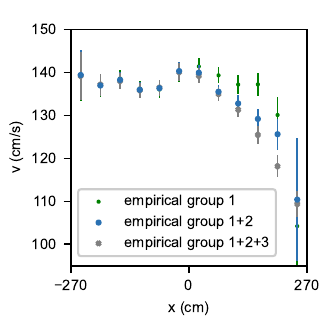}
         \caption{}
         \label{fig:emp-groups}
     \end{subfigure}
     \begin{subfigure}[t]{0.3\textwidth}
         \includegraphics[width=\textwidth]{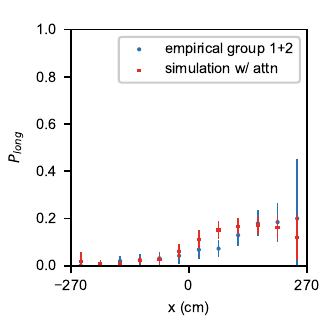}
         \caption{}
         \label{fig:prob-long}
     \end{subfigure}
     \\
     \begin{subfigure}[t]{0.3\textwidth}
         \includegraphics[width=\textwidth]{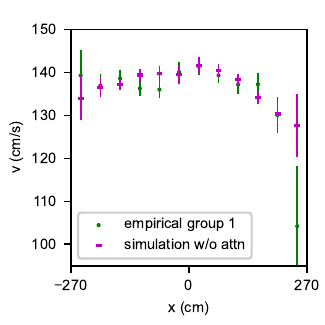}
         \caption{}
         \label{fig:v_emp_and_sim_wo_attn}
    \end{subfigure}
    \begin{subfigure}[t]{0.3\textwidth}
         \includegraphics[width=\textwidth]{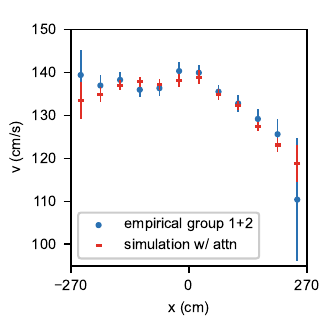}
         \caption{}
         \label{fig:v_emp_and_sim_w_attn}
     \end{subfigure}

    \caption{\textbf{Simulation results compared to empirical data}. To better visualize the contribution of our model, we split all pedestrians in the empirical dataset into 3 disjoint groups: (1) commuters; (2) pedestrians with attention-based movement behaviors; (3) consumers. (see definitions in \S \ref{sec:boundary}) (\subref{fig:emp-groups}) Mean walking speed $v$ as a function of x coordinate in empirical data. The error bars represent the 95\% CIs for the metrics. We compare the slowing effects when different groups are included in the population. (\subref{fig:prob-long}) The proportion of long-attention pedestrians $P_{long}$ as a function of x coordinate.  (\subref{fig:v_emp_and_sim_wo_attn}) Mean walking speed $v$ as a function of x coordinate for commuters (group 1). Green: empirical measurement. Magenta: simulation data with visual attention module disabled (simulation of group 1). (\subref{fig:v_emp_and_sim_w_attn}) Mean walking speed $v$ as a function of x coordinate for commuters and pedestrians with attention-based movement (group 1+2). Blue: empirical measurement. Red: simulation data with visual attention module enabled (simulation of group 1+2). 
    }
    \label{fig:sim-results}
\end{figure}

We firstly show that attention-based movement pedestrians make a major contribution to the slowing effects. In Fig. \ref{fig:emp-groups}, the difference between the mean walking speed function of all pedestrians (group 1+2+3, see group definitions in \S \ref{sec:boundary}) and that of group 1+2, is smaller than the difference between group 1+2+3 and group 1. It shows the removal of group 3 (consumers) does not impose a major impact to pedestrian flows, especially for x coordinate intervals where most pedestrians appear. But group 2 does so, which makes itself as our focus in this paper.

Our simulation was generally in accordance with the empirical data. In Fig.~\ref{fig:prob-long}, the proportion of long-attention pedestrians $P_{long}$ decreased with the distance to the store entrance. In Fig.~\ref{fig:v_emp_and_sim_w_attn}, walking speed function $v$ showed a decaying slowing effect of the store when the distance increases.

We also show our simulation results without visual attention module in Fig.~\ref{fig:v_emp_and_sim_wo_attn} (in magenta), where we observe the walking speed distribution aligns with empirical data where we only keep commuters in the empirical data (in green). After we include pedestrians with visual attention into the empirical data (in blue, see Fig.~\ref{fig:v_emp_and_sim_w_attn}), we observe the slowing effect that can only be captured when we enable the visual attention module in our simulation (in red). 
Therefore, we prove that the slowing effect stems from visual attention module and its regulation on locomotion, rather than the over-fitting from SFM or boundary condition parameters.

\section{Case Study}
\label{sec:5}

In this section, we explore how architectural design features may influence pedestrian efficiency and retail potential using our model. The first application aims to reveal the influence of three key design feature variables (corridor width, store entrance width, and display depth) qualitatively. The second one shows the process of striking a balance between efficiency and retail potential by changing the corridor width incrementally in simulations.

\subsection{Qualitative Analysis on Design Feature Impacts} \label{sec:case1}

\begin{figure}[htbp]
    \centering
    \begin{adjustbox}{max width=\textwidth}
     \begin{tabular}{cccc}
     \includegraphics[width=.2\linewidth]{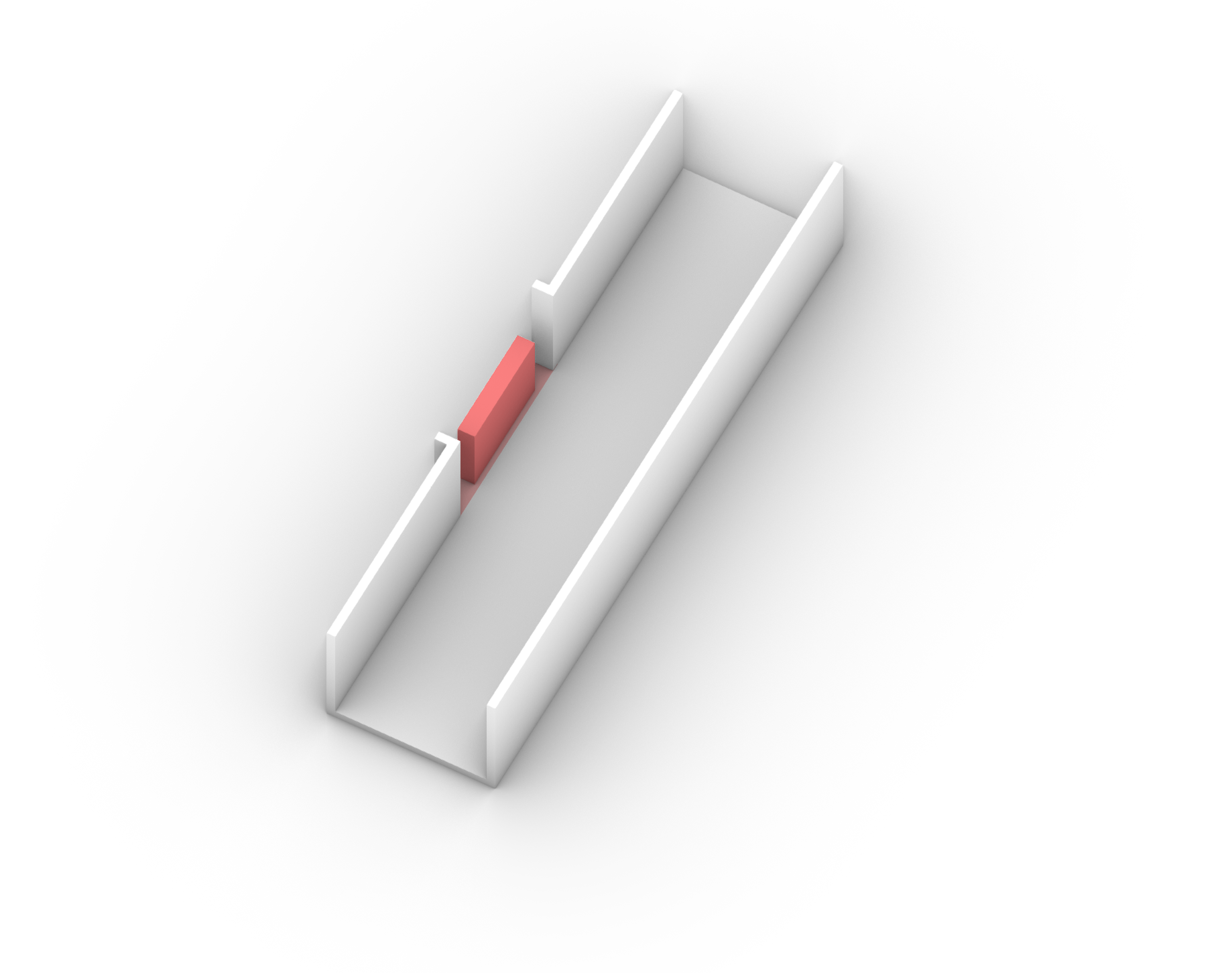} & \includegraphics[width=.2\textwidth]{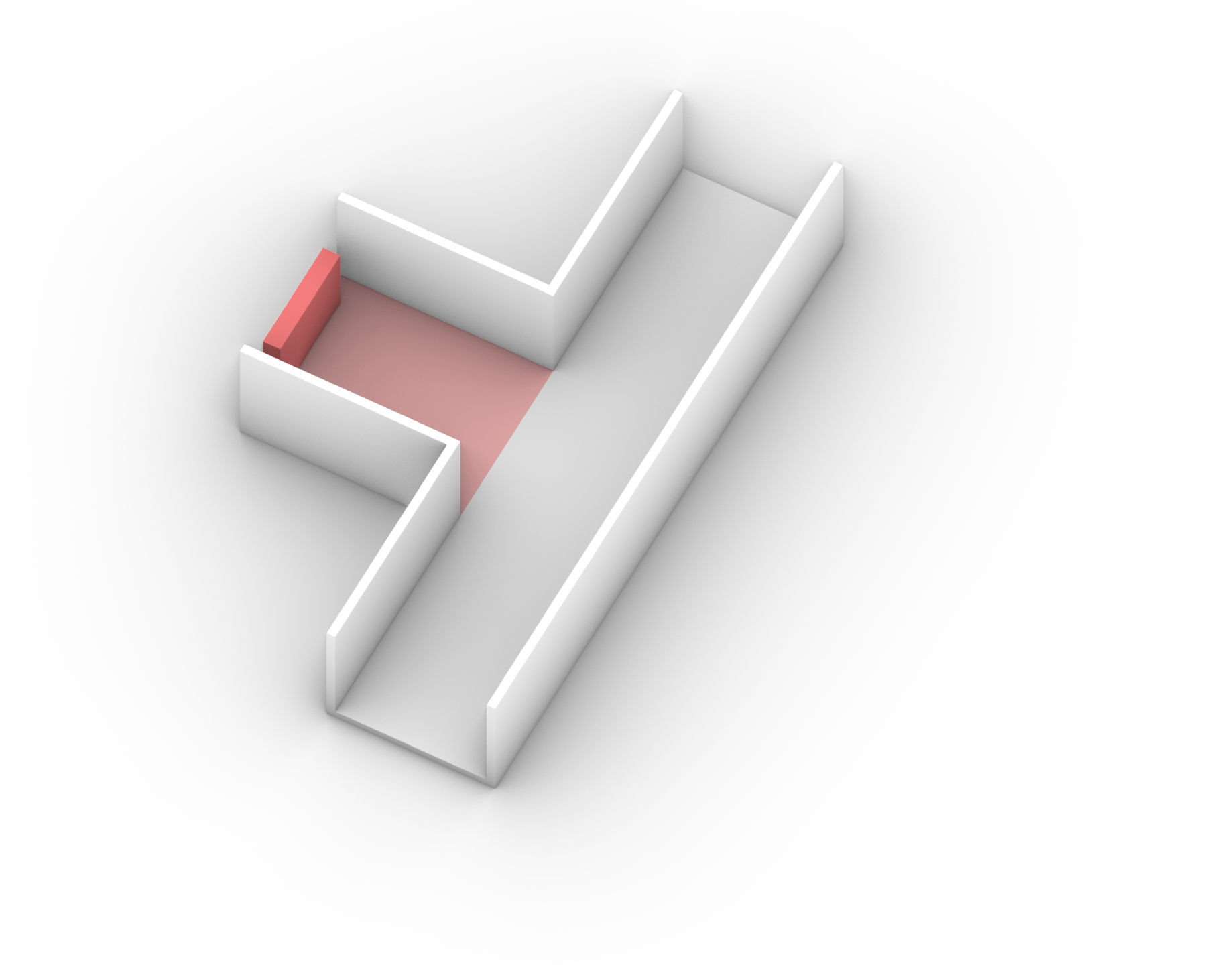} & \includegraphics[width=.2\textwidth]{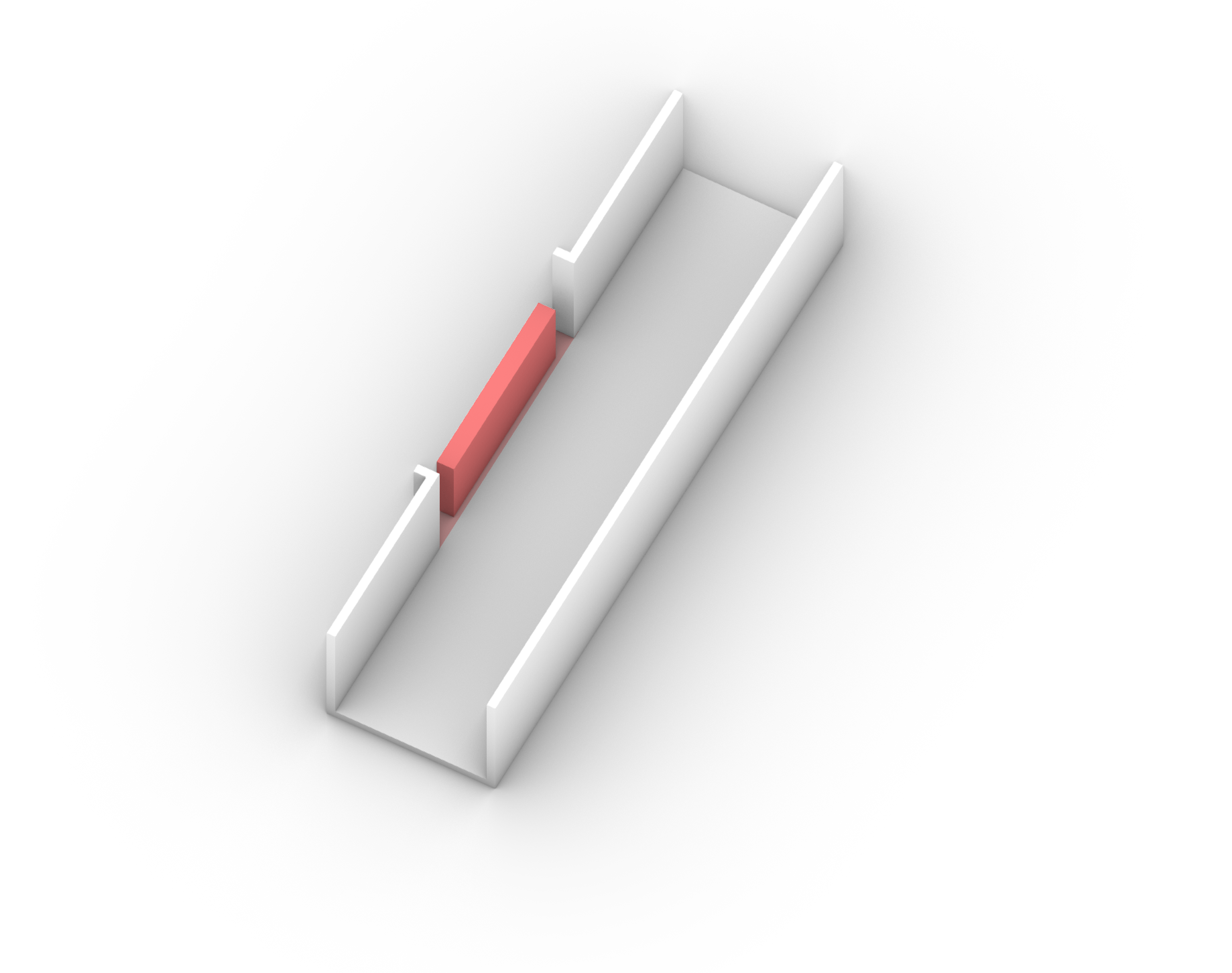} & \includegraphics[width=.2\textwidth]{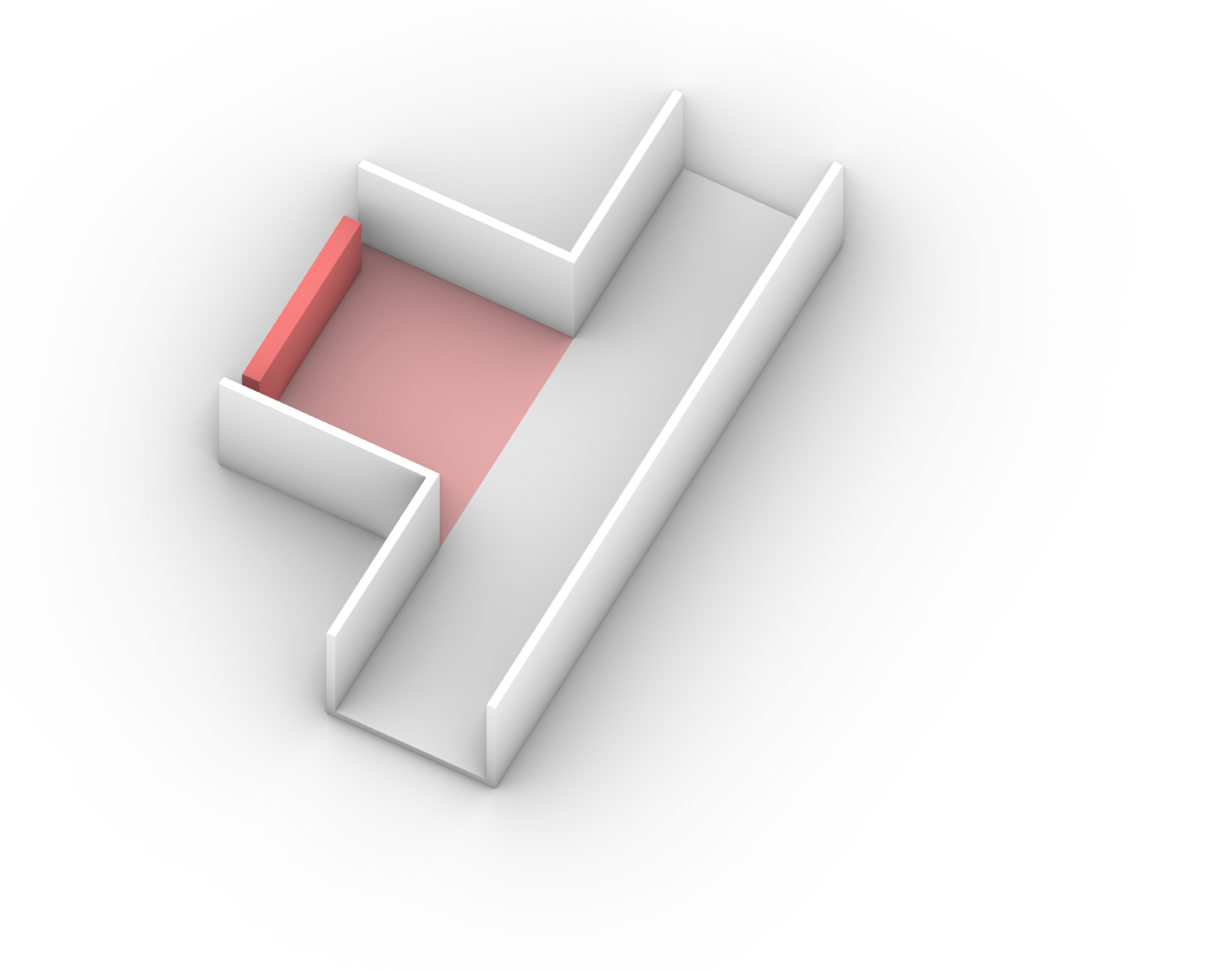}\\
     $L_c = 3.5, L_s = 4.2, L_d=0.5$ &$L_c = 3.5, L_s = 4.2, L_d=5$ &$L_c = 3.5, L_s = 6, L_d=0.5$ & $L_c = 3.5, L_s = 6, L_d=5$\\[2ex]
     \includegraphics[width=.2\textwidth]{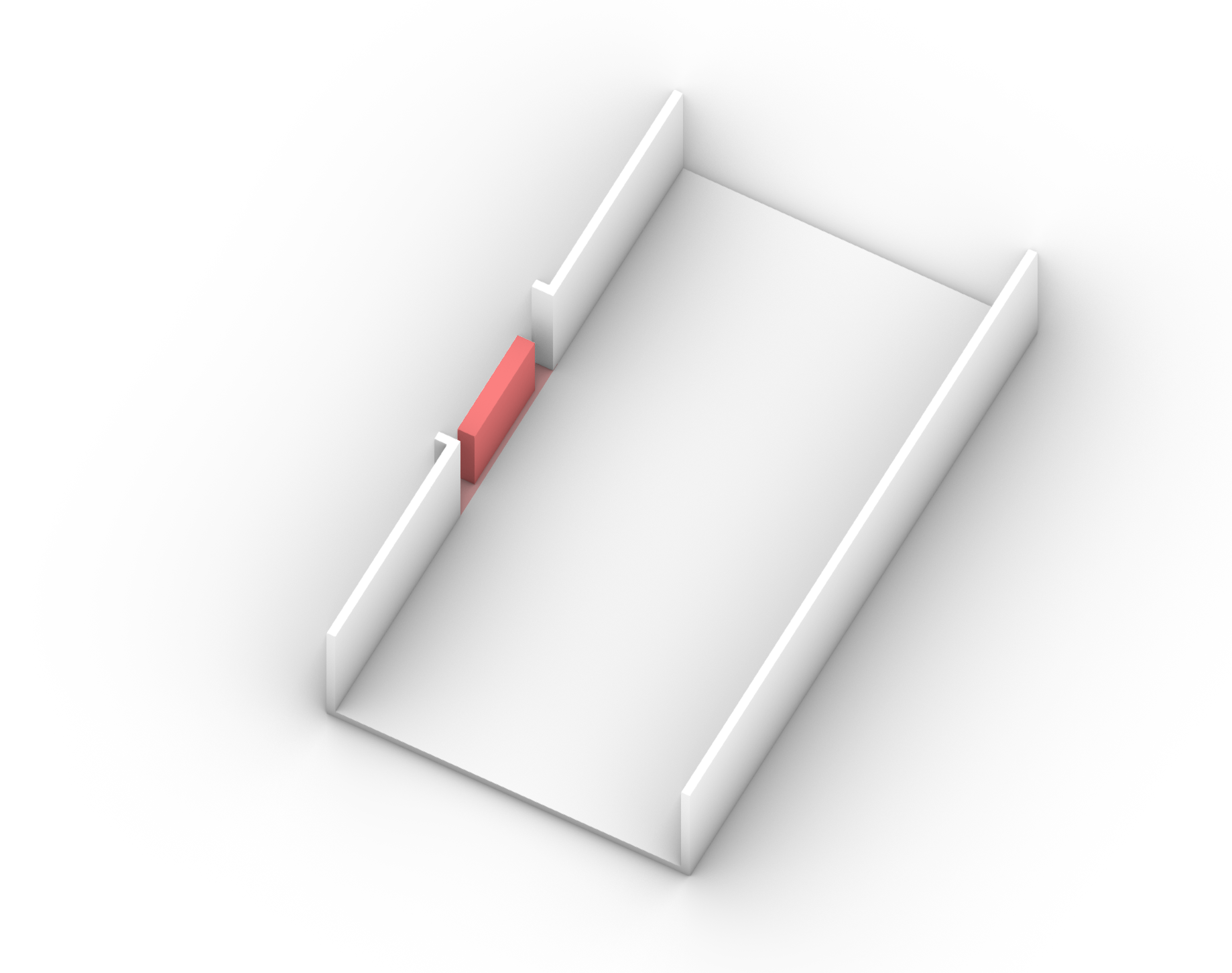} & \includegraphics[width=.2\textwidth]{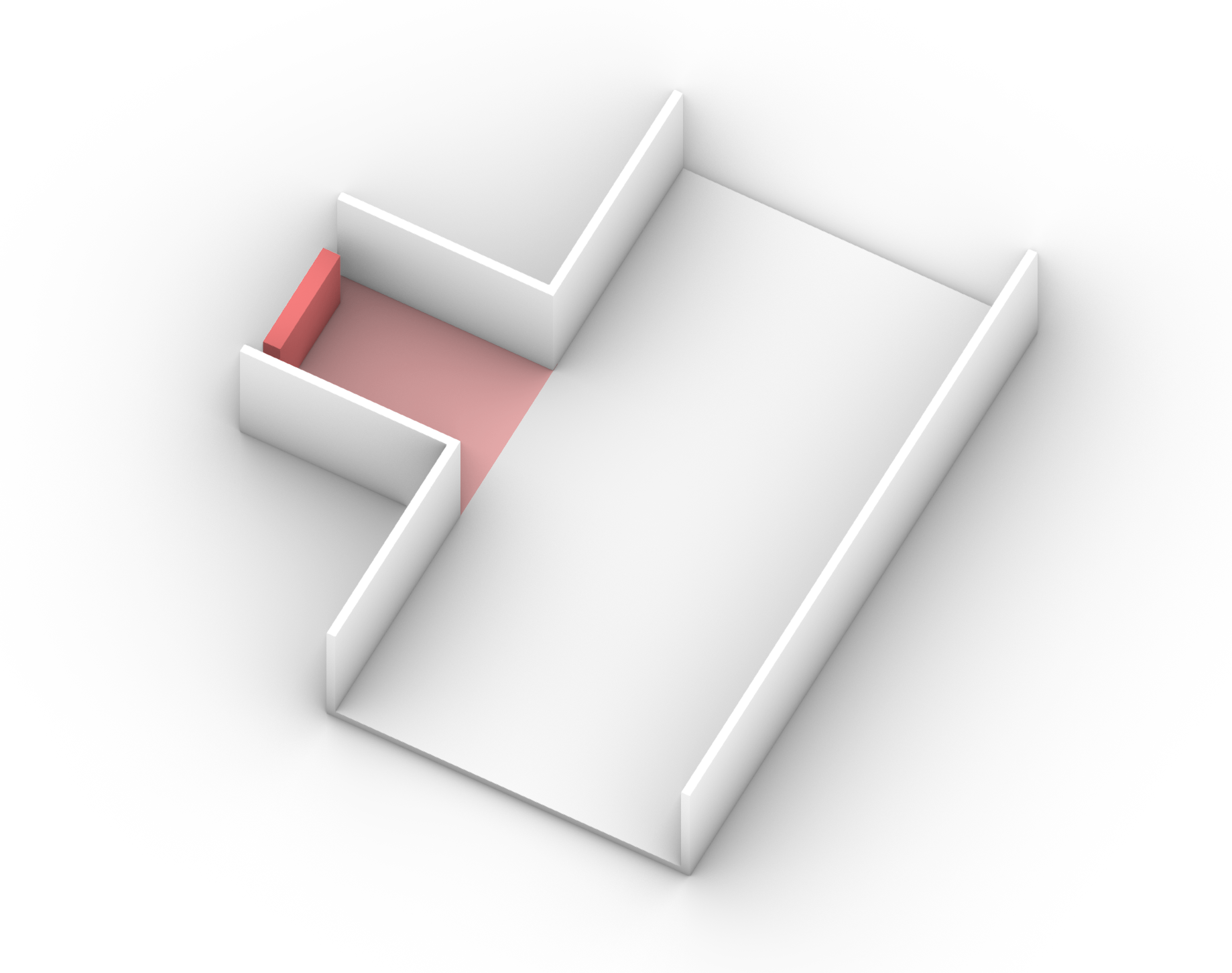} & \includegraphics[width=.2\textwidth]{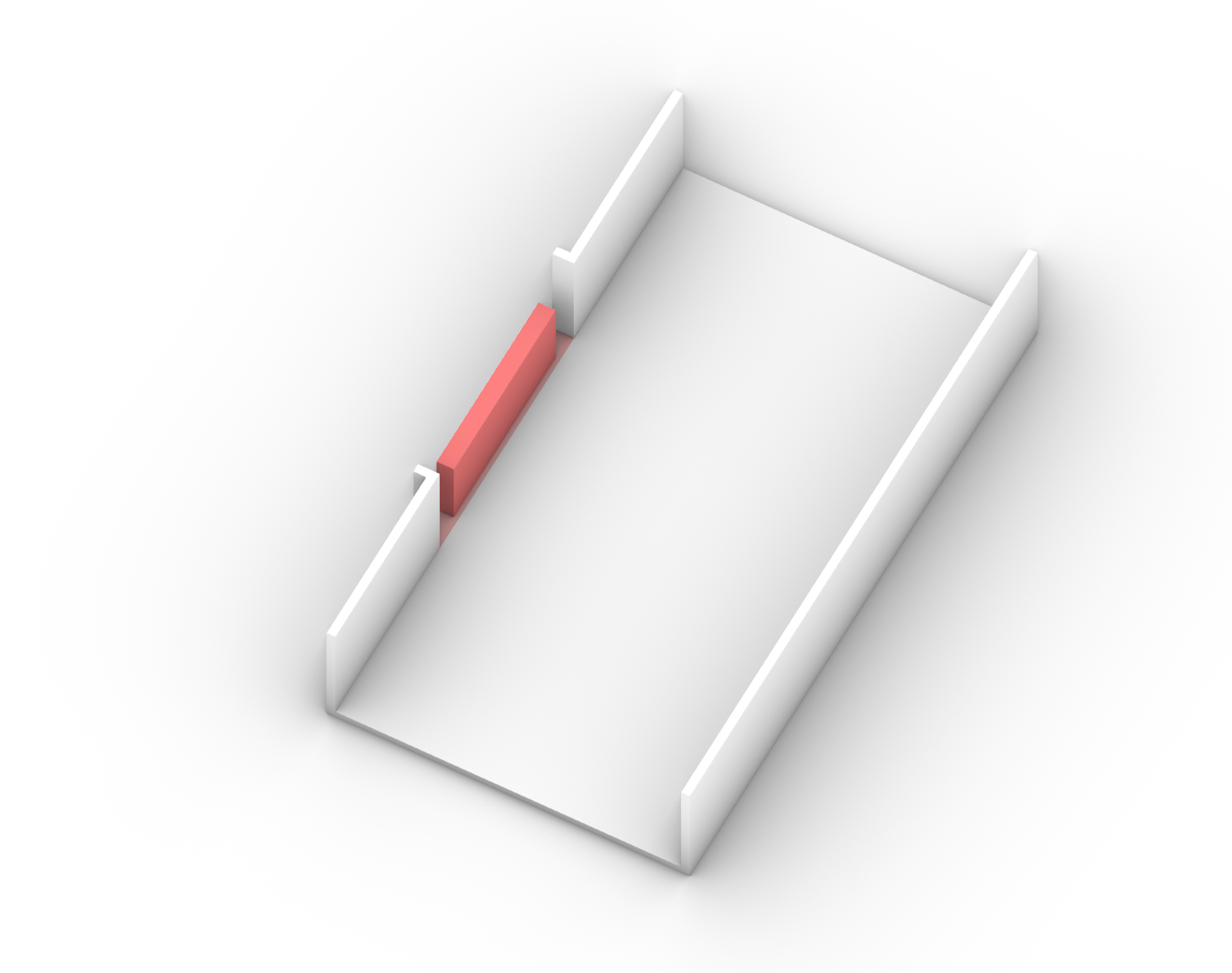} & \includegraphics[width=.2\textwidth]{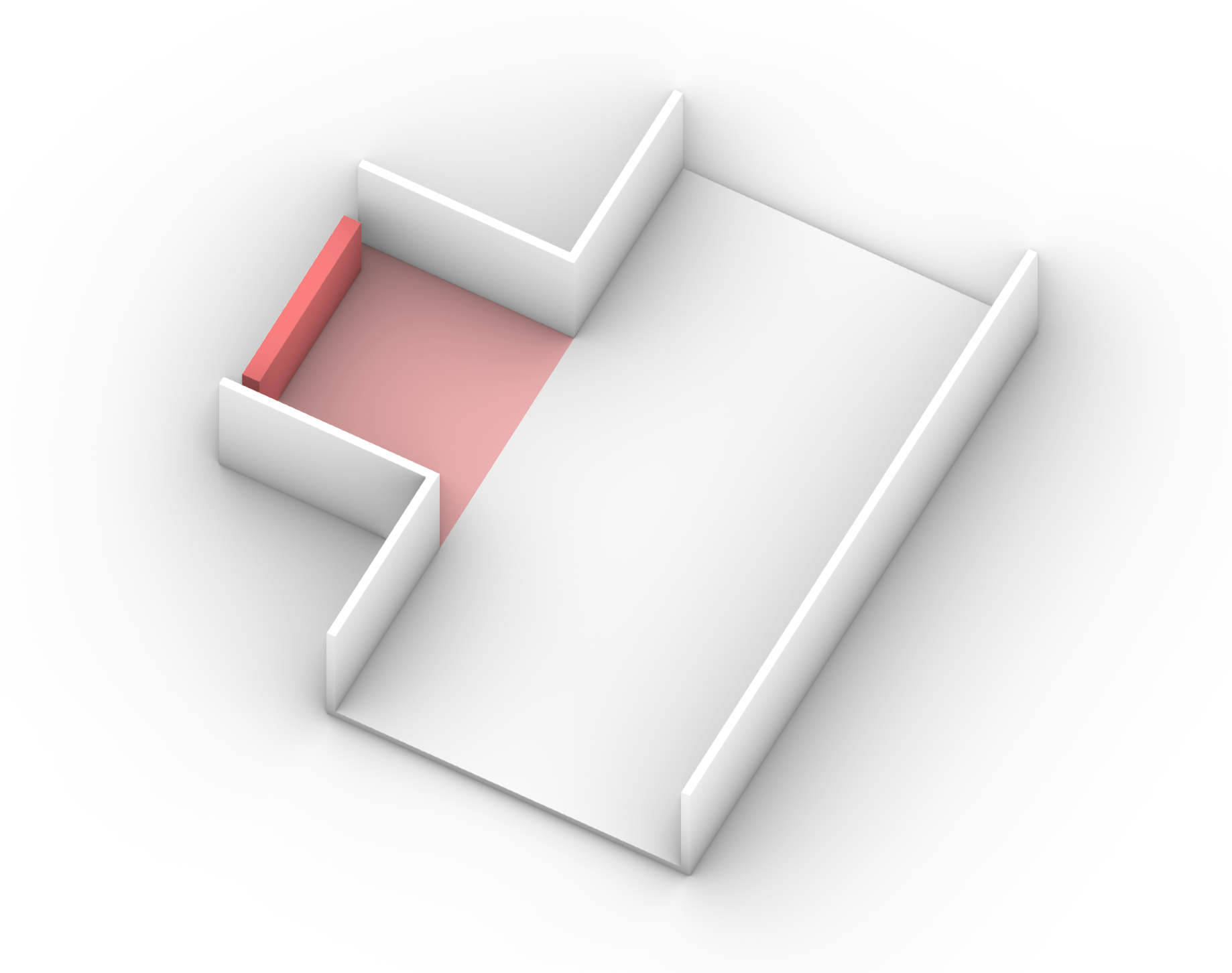}\\
     $L_c = 8, L_s = 4.2, L_d=0.5$ &$L_c = 8, L_s = 4.2, L_d=5$ &$L_c = 8, L_s = 6, L_d=0.5$ & $L_c = 8, L_s = 6, L_d=5$
     \end{tabular}
     \end{adjustbox}

    \caption{\textbf{Design settings with varying features in isometric views}. White floors represent the corridor segment. Light red floors are store areas. And the red volumes represent store displays. Corridor width $L_c$, store width $L_s$, and store depth $L_d$ are all in the unit of meter here. }
    \label{fig:factorial-floor-plan}
\end{figure}

To reduce the computation complexity in simulation, discretizations are applied to design feature variables: We pick two values for each variable. Then we compare the simulation results between all combinations of the values. The values are discretized as: corridor width $L_c = \{3.5m, 8m \}$, store entrance width $L_s = \{4.2m, 6m \}$, and display depth (defined in \S \ref{sec: angular}) $L_d = \{0.5m, 5m\}$. The combination of these values provide 8 test conditions (illustrated in Fig. \ref{fig:factorial-floor-plan}).

The pedestrian flows are generated in two directions. For each direction, the flow rate for a unit of corridor width is fixed to $0.08$ $people \cdot m^{-1} \cdot s^{-1}$. Other parameters are set as Tab. \ref{tab: sim_config}, \ref{subtab:locomotion-regulation}, \ref{subtab:sfm}, and \ref{subtab:boundary}.

In the analysis, the average speed loss $\Delta v$ and long-attention-pedestrian proportion $P_{long}$ are compared between trials. The speed loss here is defined as the speed difference between the simulation and an ideal simulation, which has the same corridor width but has no stores. The two metrics above are shown in their spatial distribution form. In other words, rather than analyzing the corridor as a whole, the corridor is divided into rectangular cells. And the metrics are calculated for each cell using the pedestrian trajectory segments inside it.

\begin{figure}
    \centering
    \begin{subfigure}{.45\textwidth}
         \centering
         \includegraphics[width=\linewidth]{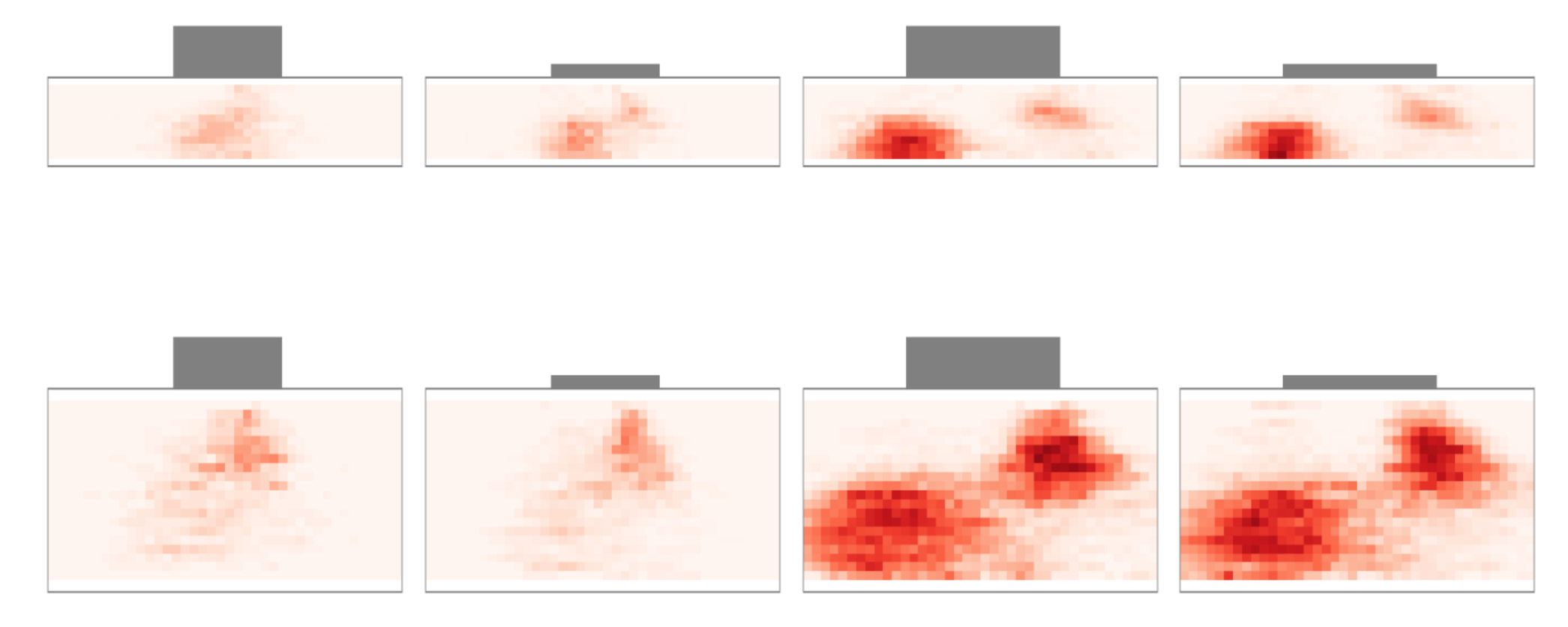}
         \caption{}\label{subfig:factorial_long}
     \end{subfigure}
    \begin{subfigure}{.45\textwidth}
         \centering
         \includegraphics[width=\linewidth]{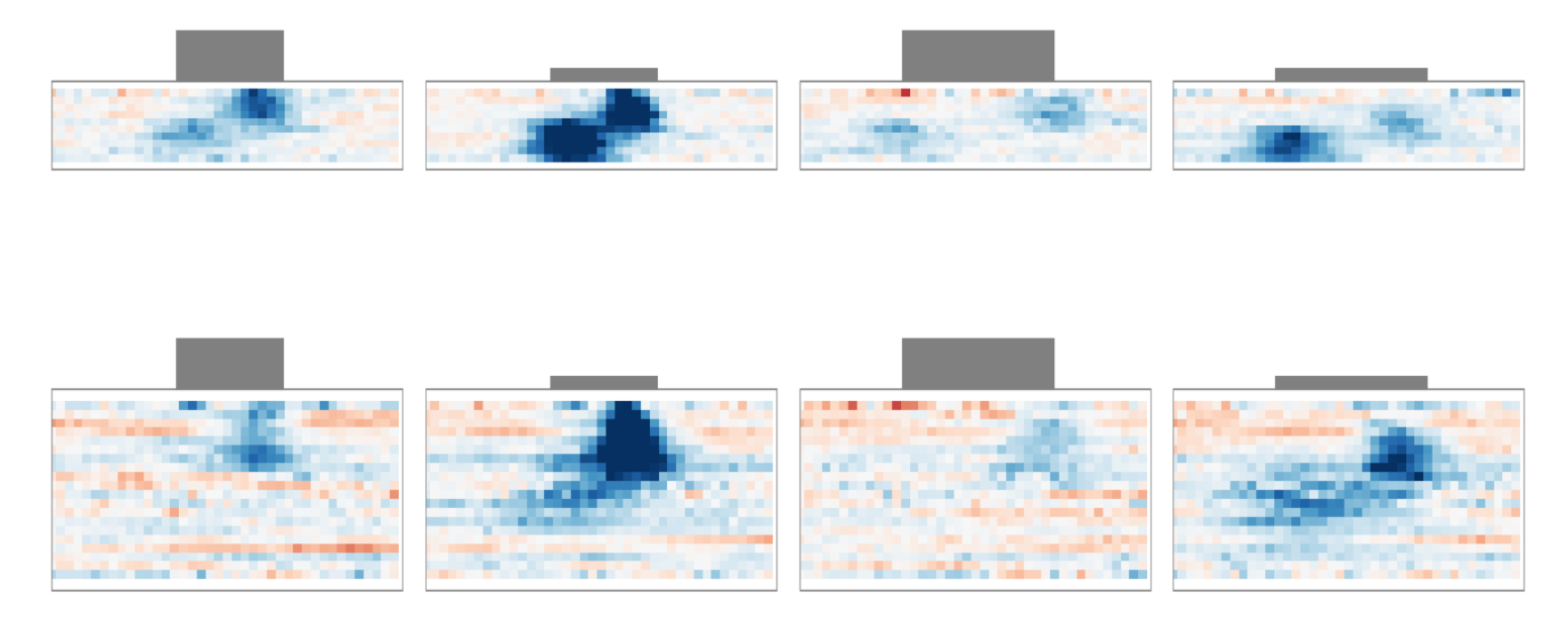}
         \caption{}\label{subfig:factorial_delta}
     \end{subfigure}

    \caption{\textbf{Simulations with varying design features} Refer to Fig. \ref{fig:factorial-floor-plan} for corresponding 3D representation. (\subref{subfig:factorial_long}) The distribution map of $P_{long}$, which means the probability of observing a pedestrian with long-time visual attention to the store; (\subref{subfig:factorial_delta}) The distribution map of $\Delta v$, which means the average speed loss compared to the same corridor without a store.  Maps are top-down views of the corridor in real-world dimensions. The grey boxes represent the dimensions of the store width and display depth. }
    \label{fig:factorial}
\end{figure}

As illustrated in Fig.~\ref{subfig:factorial_long}, design feature variables imposed complex impacts on the distribution of $P_{long}$. $P_{long}$ did not always decrease with the distance to the store. The trend was conditioned on corridor widths. For very narrow corridors, a larger proportion of pedestrians drew their attention to the store when they were located further away from it.
This can be attributed to the constrained visual coverage in narrow corridors. For pedestrians who walk on the store side, they will suffer a larger observation angle in a narrow corridor, which discourages them to draw their attention for a long time. For the other two variables, store widths were positively correlated to visual attention while display depths played little role.

While not influential to $P_{long}$, display depths greatly affected $\Delta v$. As shown in Fig.~\ref{subfig:factorial_delta}, shallower storefront displays led to larger speed decreases, which resulted from the angular speed adaptations. Since store displays are much closer, pedestrians have to decrease speed more to keep the angular speed at a low level. We also observed such impacts interacting with the other two variables. For instance, the speed decrease got smaller when the store became wider. That can be explained by the spread-out of visual attention distribution, which makes pedestrians get attracted when they are further away from the store, lowering their need to maintain the angular speed.

\subsection{Optimization: Strike the Balance} \label{sec:case2}

Finally, to demonstrate our model can help designers find the balance between efficiency and retail potential, we used our model to run a local sensitivity analysis on corridor width in two store display depth settings. These options resemble decisions in real practice where designers need to find an ideal facility dimension when other external factors are fixed.

\begin{figure}[htbp]
    \centering
    \begin{adjustbox}{max width=\textwidth}
     \begin{tabular}{ccccc}
     \includegraphics[width=.15\linewidth]{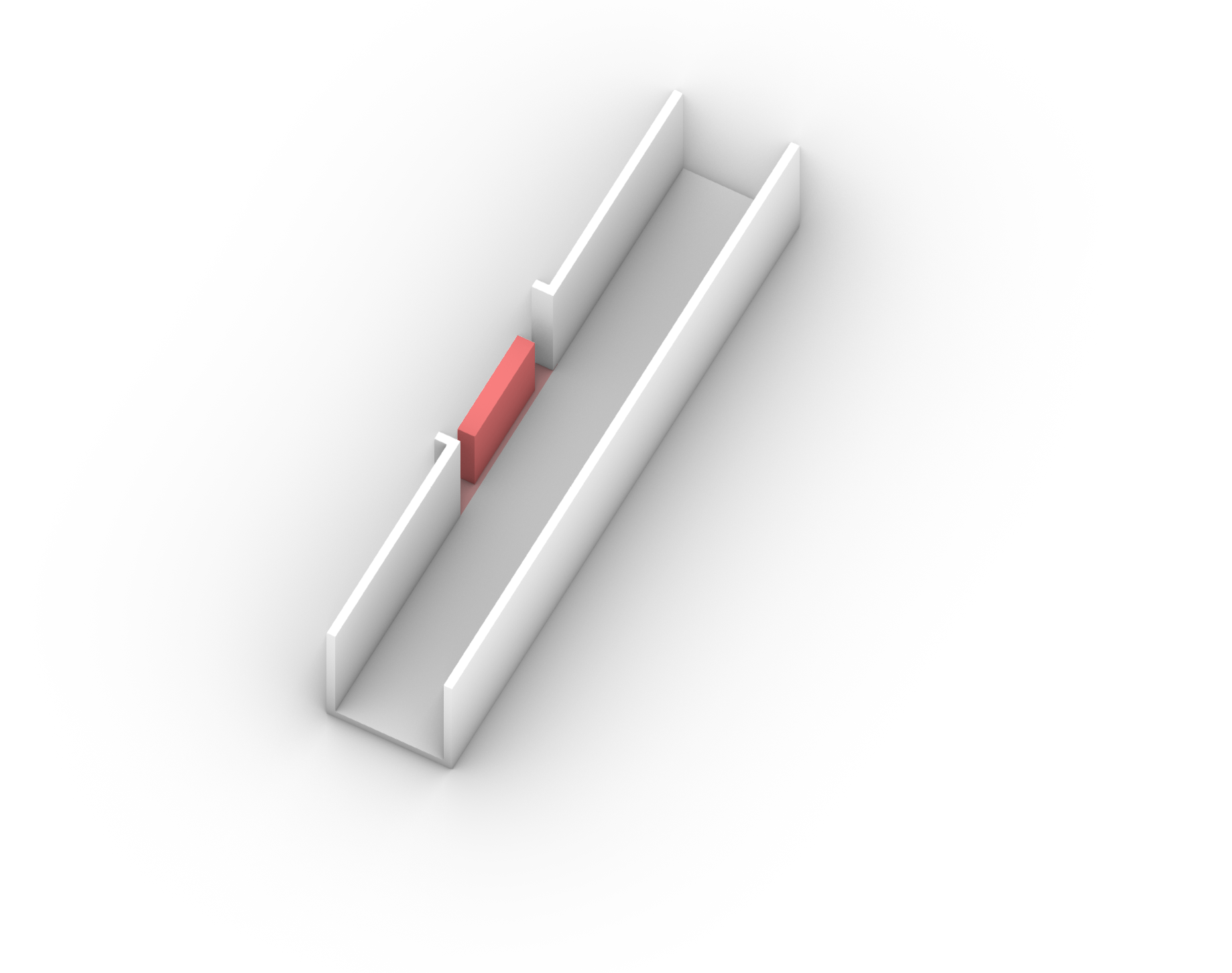} & \includegraphics[width=.15\linewidth]{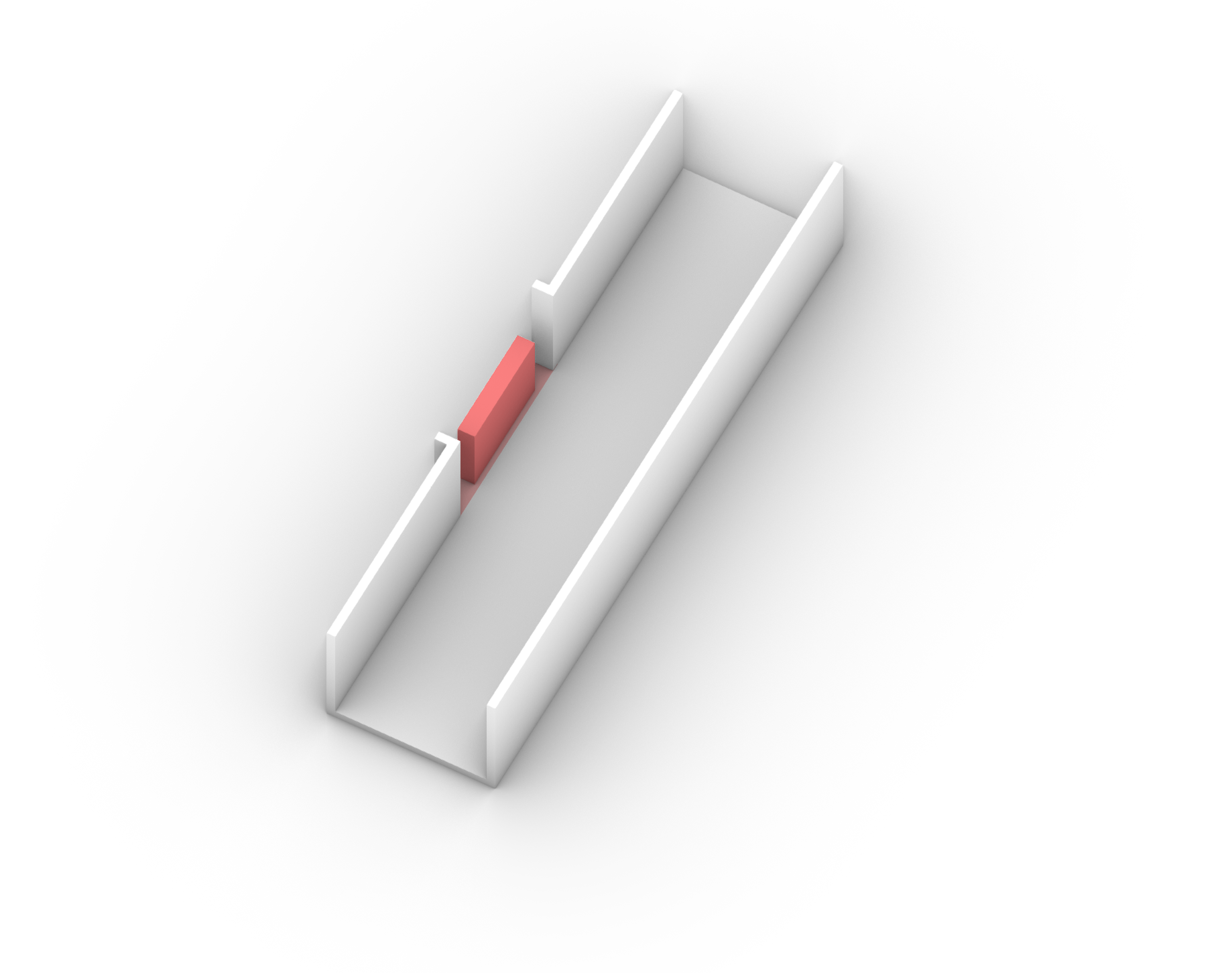} & \includegraphics[width=.15\linewidth]{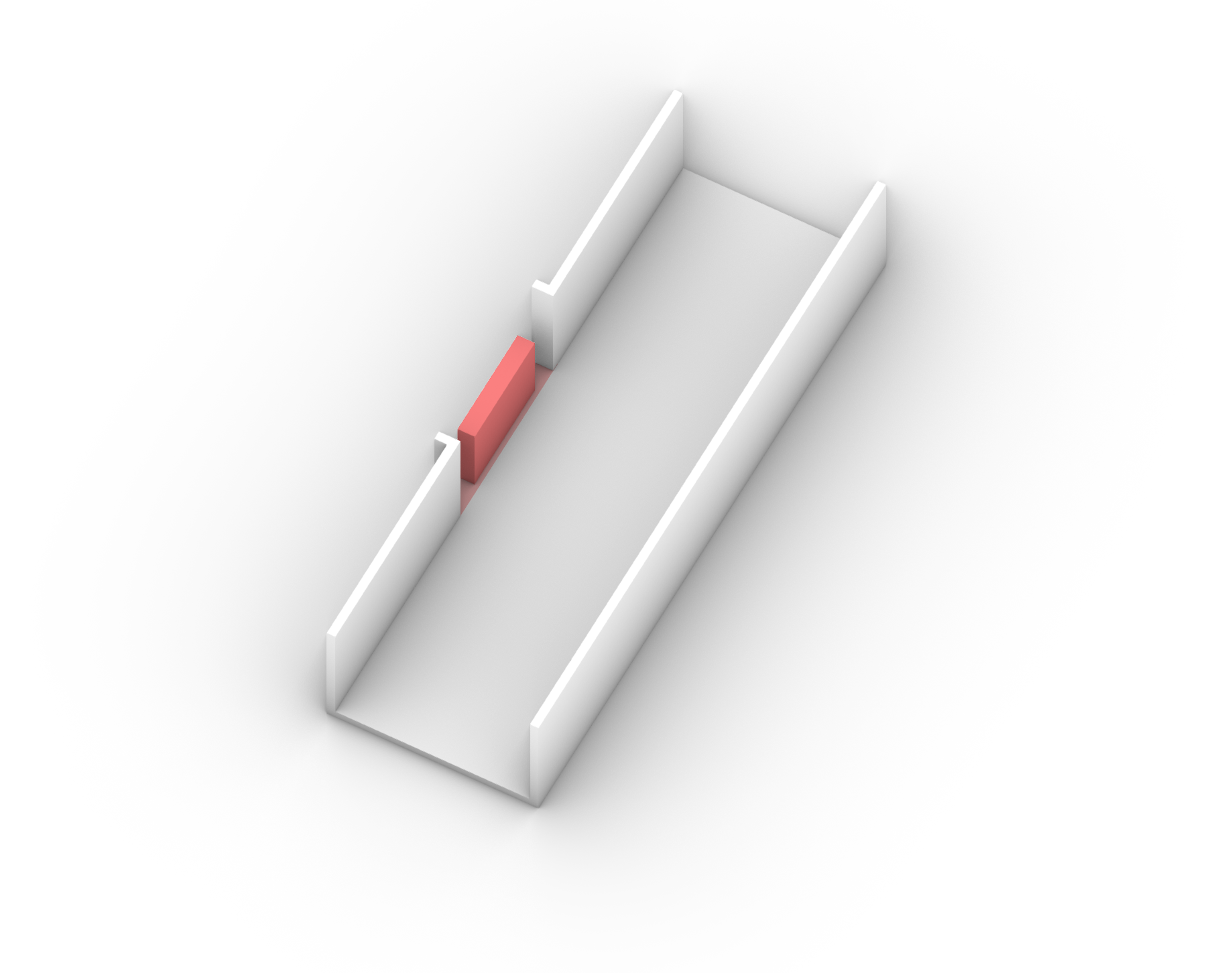} & \includegraphics[width=.15\linewidth]{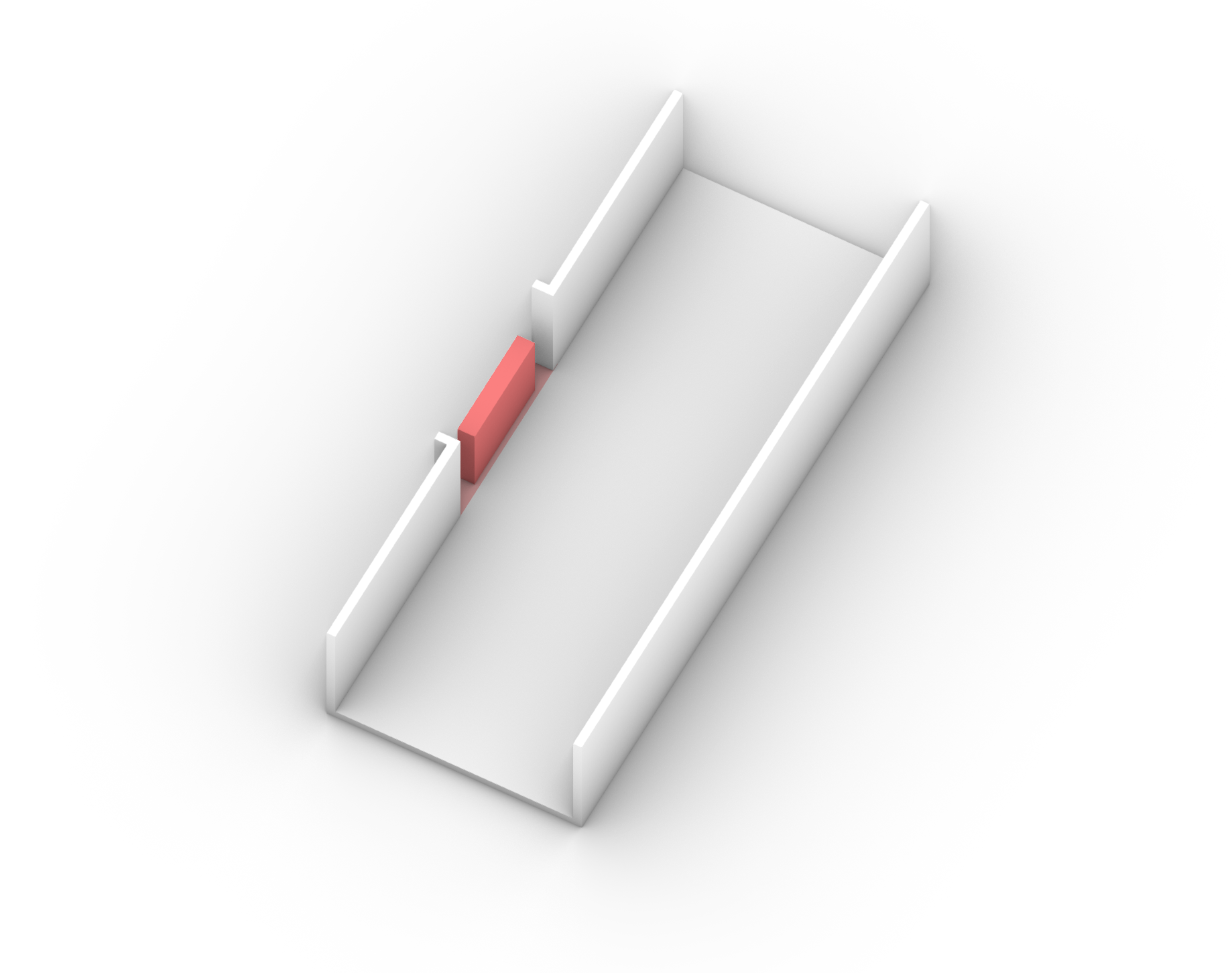} & \includegraphics[width=.15\linewidth]{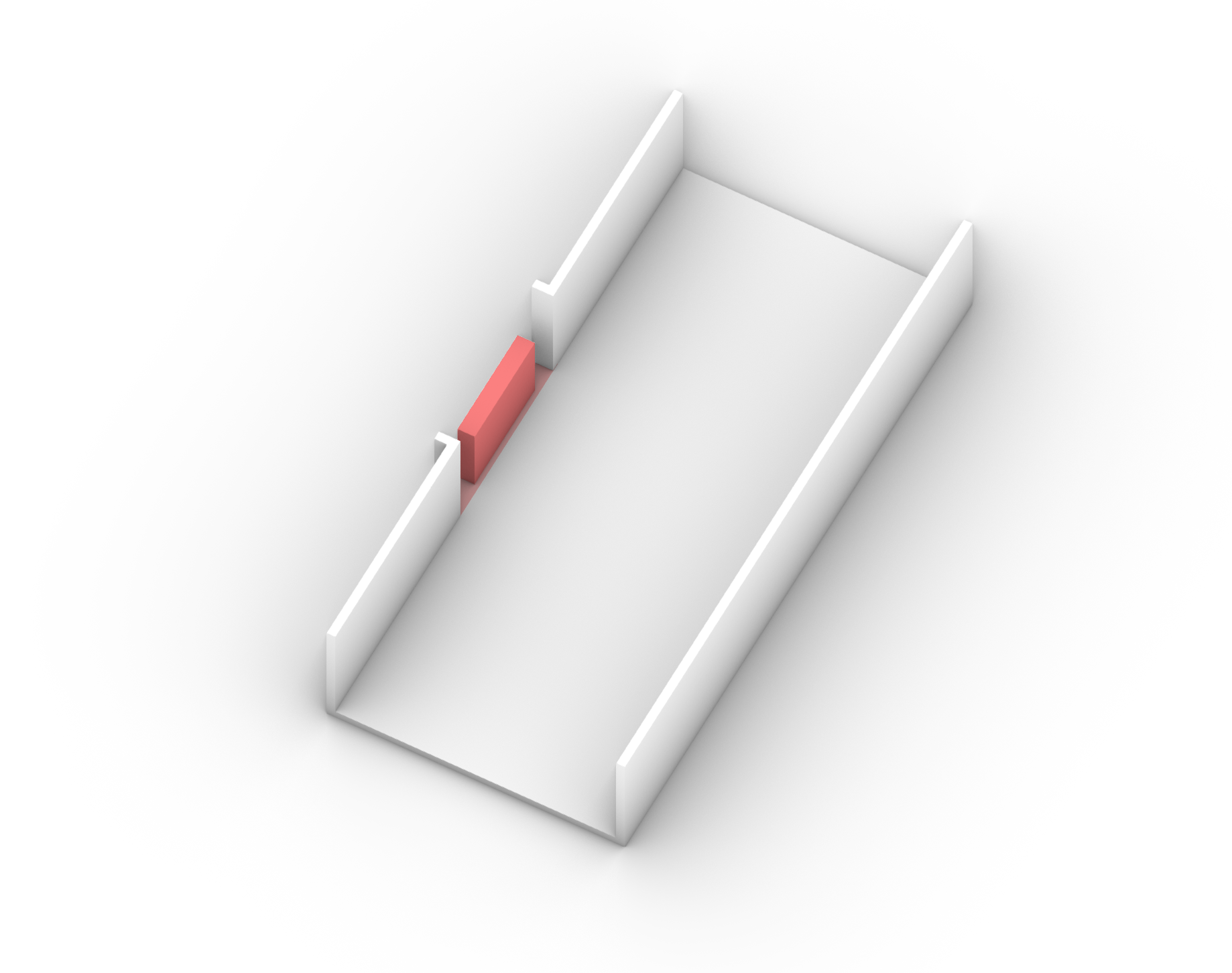} \\
     $L_c = 2.5m, L_d=0.5m$ &$L_c = 3.5m, L_d=0.5m$ &$L_c = 4.5m, L_d=0.5m$ & $L_c = 5.5m, L_d=0.5m$ & $L_c = 6.5m, L_d=0.5m$\\[2ex]
     \includegraphics[width=.15\linewidth]{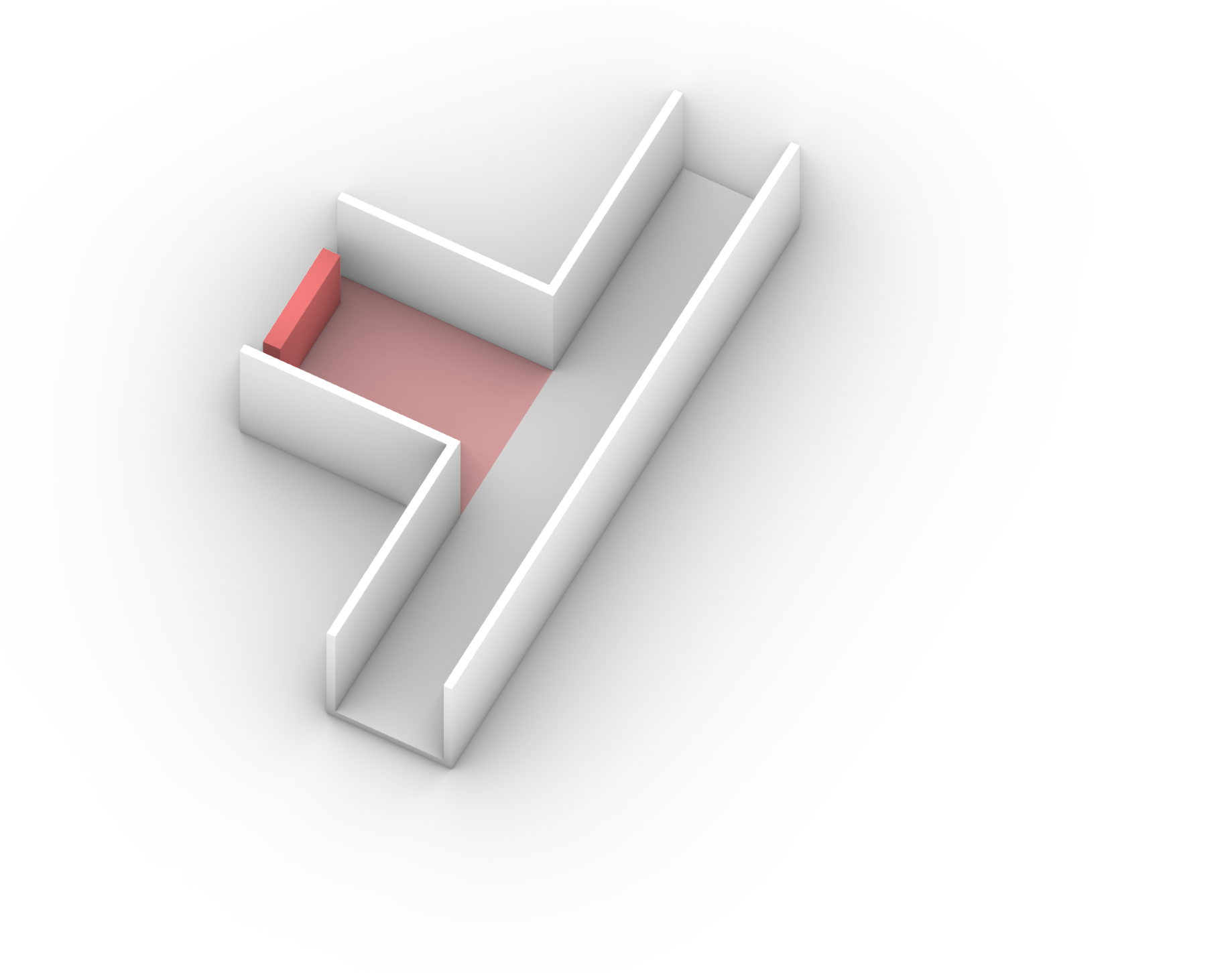} & \includegraphics[width=.15\textwidth]{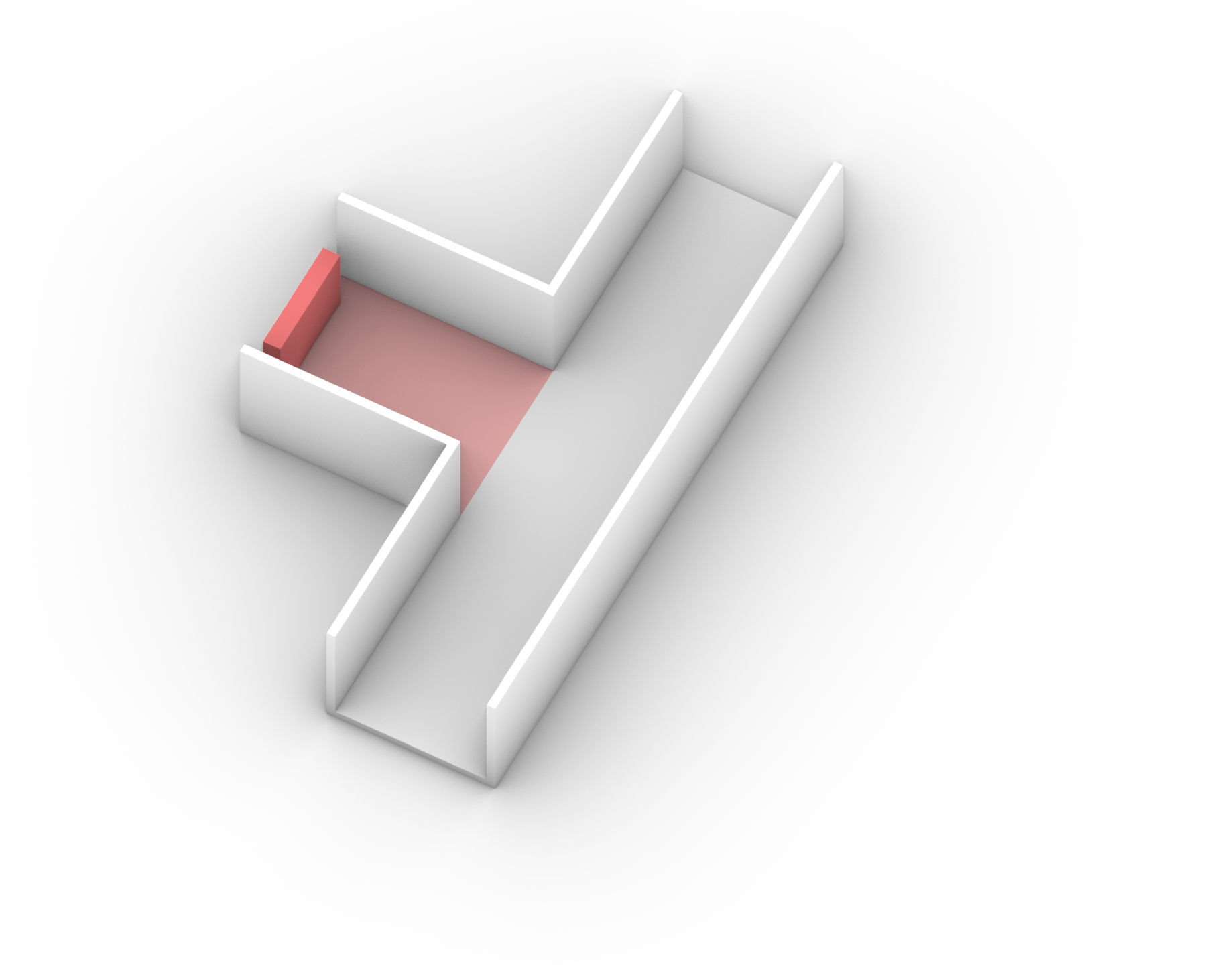} & \includegraphics[width=.15\textwidth]{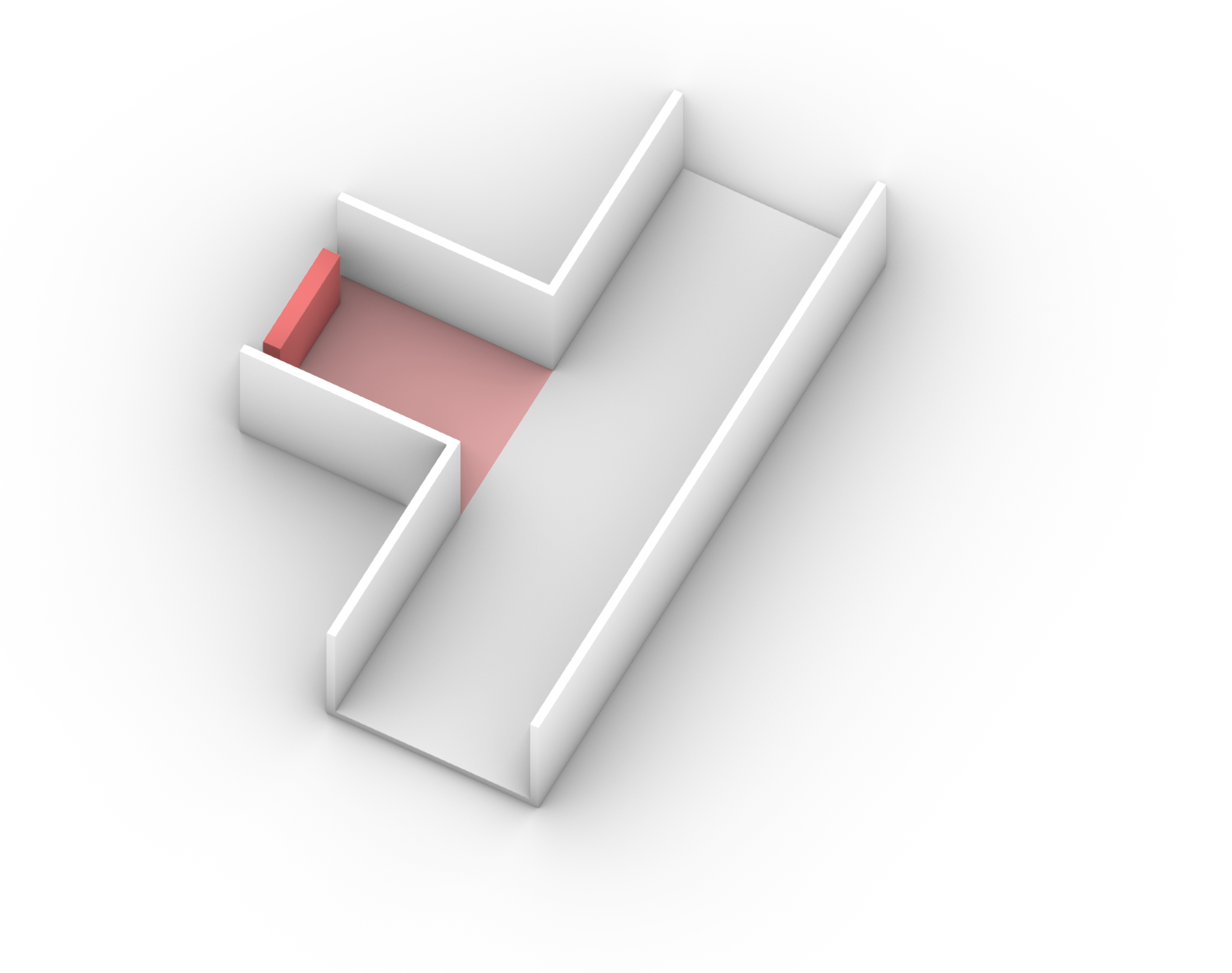} & \includegraphics[width=.15\textwidth]{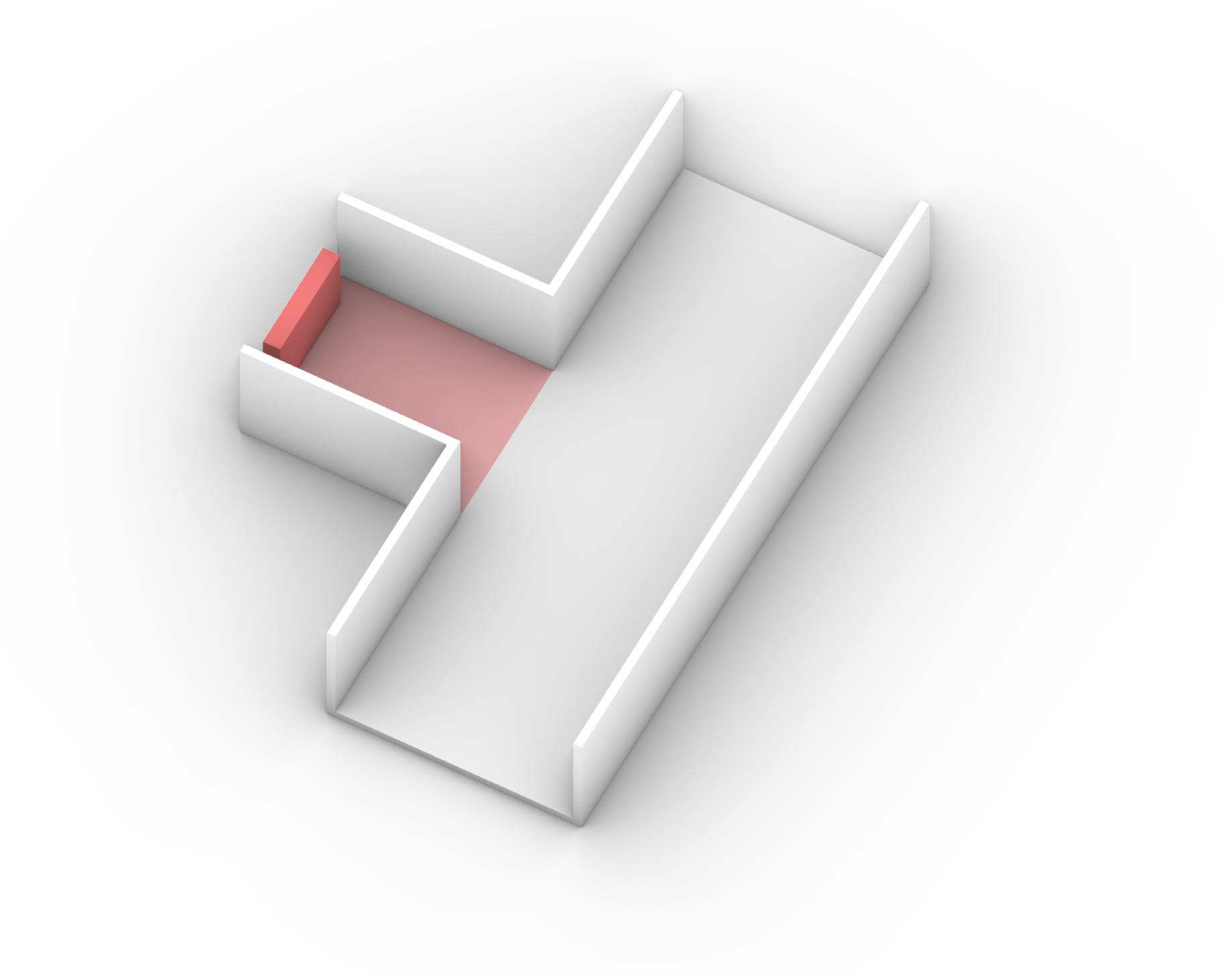} & \includegraphics[width=.15\textwidth]{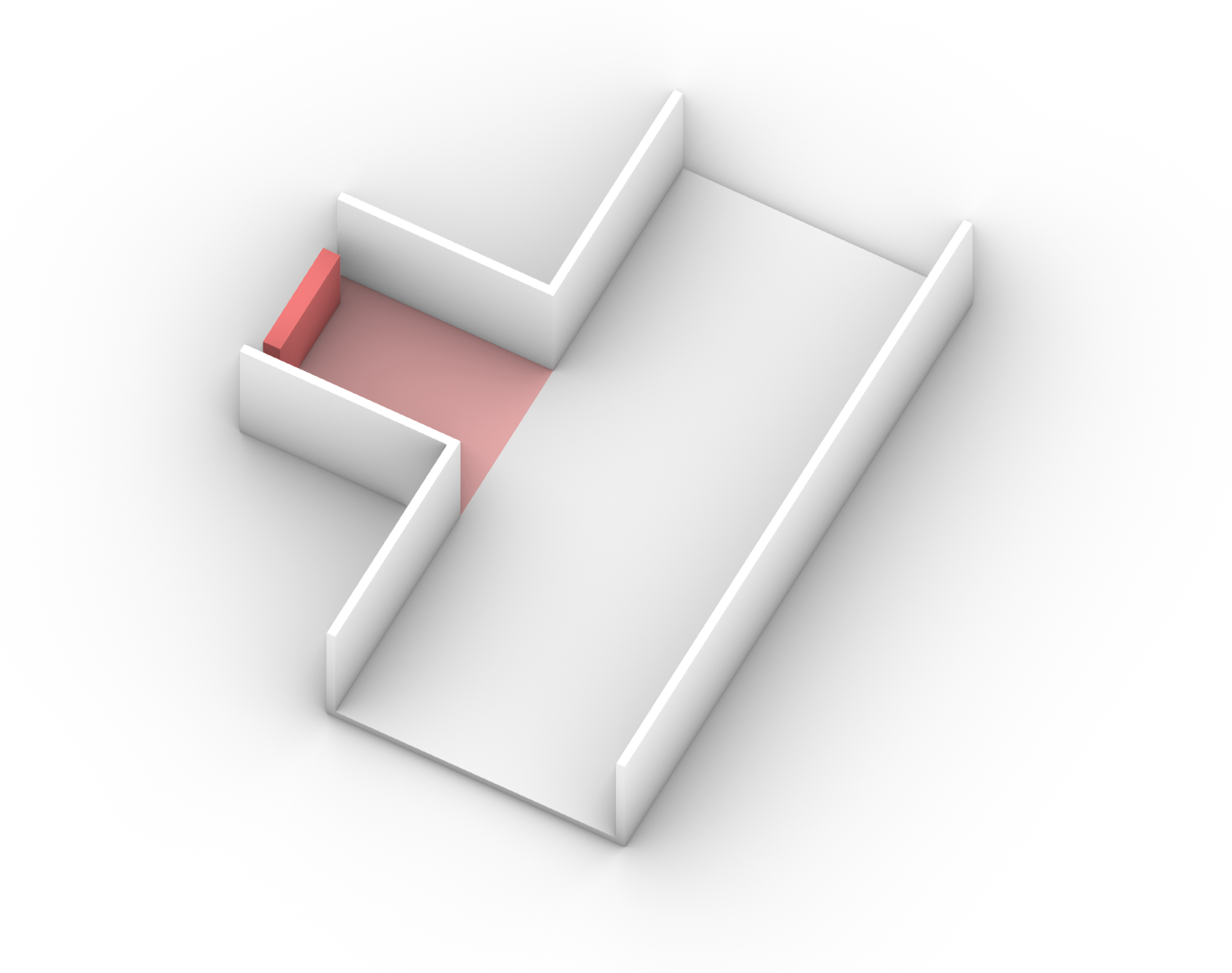} \\
     $L_c = 2.5m, L_d=5m$ &$L_c = 3.5m, L_d=5m$ &$L_c = 4.5m, L_d=5m$ & $L_c = 5.5m, L_d=5m$ & $L_c = 6.5m, L_d=5m$\\
     \end{tabular}
     \end{adjustbox}

    \caption{\textbf{Design settings with varying corridor width in isometric views}. White floors represent the corridor segment. Light red floors are store areas. And the red volumes represent store displays. }
    \label{fig:sa-floor-plan}
\end{figure}

In the simulation, display depths are set as 5m and 0.5m respectively. In both cases, the flow rates for each direction are fixed to 2 people/s, and the entrance widths are fixed to 4.2m. Other parameters are set as Tab. \ref{tab: sim_config}, \ref{subtab:locomotion-regulation}, \ref{subtab:sfm}, and \ref{subtab:boundary}. We incrementally change the corridor width (illustrated in Fig. \ref{fig:sa-floor-plan}) to evaluate its impact on two metrics (long-attention pedestrian proportion $P_{long}$ and average walking speed $v$). Here the corridor is analyzed as a whole without spatial discretizations.

\begin{figure}[htbp]
     \centering
     \begin{subfigure}[t]{0.3\textwidth}
         \includegraphics[width=\textwidth]{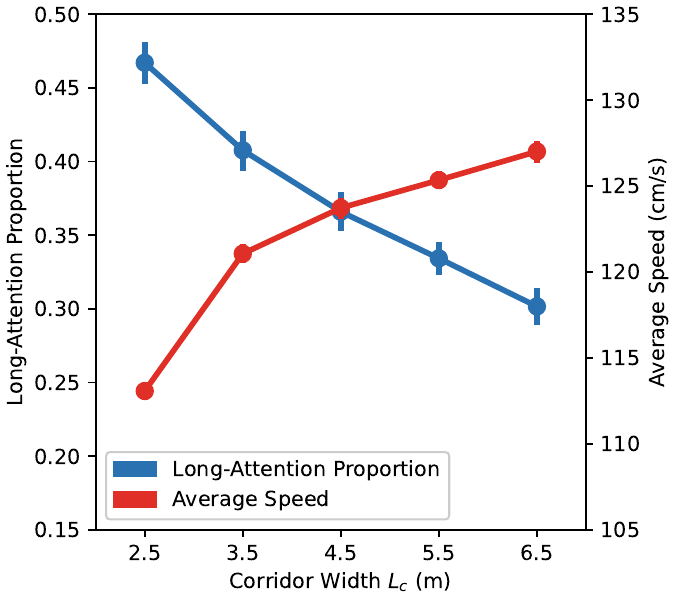}
         \caption{}
     \end{subfigure}
     \begin{subfigure}[t]{0.3\textwidth}
         \includegraphics[width=\textwidth]{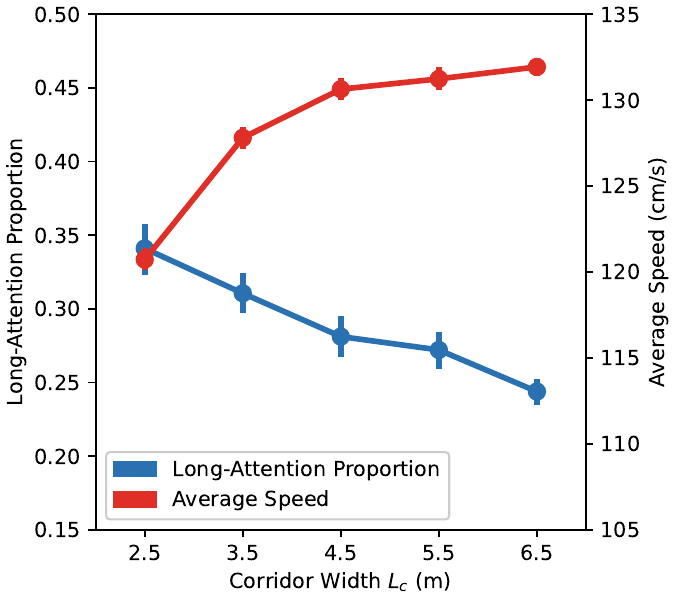}
         \caption{}
     \end{subfigure}

    \caption{\textbf{Simulations with varying corridor widths}. The proportion of long-attention pedestrians and average walking speed for all pedestrians are plotted in the same diagram with different scales. There are two scenarios where (a) store display depth is 0.5m, and (b) depth is 5m. The error bars represent the 95\% CIs for the metrics. }
    \label{fig:sim-vary-width}
\end{figure}

As shown in Fig.~\ref{fig:sim-vary-width}, we observed the trade-offs between pedestrian flow efficiency (average walking speed) and retail potential (long-attention pedestrian) in both scenarios. Although the actual balance point depends on the designers' judgments, 
the figures help designers to make data-informed decisions by providing the possible outcomes in different scenarios.

\section{Conclusion}
\label{sec:6}

In this paper, we propose a novel pedestrian simulation framework focused on providing feedback on architectural design decisions to increase retail potential and pedestrian flow efficiency for common retail spaces similar to transportation hubs. Our model uses visual attention to introduce the modification of the Social Force Models desired velocity coefficient for the goal force. The model that combines visual attention and locomotion, is compared to prior works by fitting to an existing dataset. Finally, we demonstrate qualitative results through a simple case study of a corridor and a single storefront. 

In general, the model evaluates retail potential and pedestrian flow efficiency by jointly predicting the visual attention and locomotion of the pedestrians at the same time. The visual attention, which is the proxy to retail potential, is able to simulate the transitions between the dual states of visual attention, making it adaptable to multi-store simulations. As for the model representation, we extend the model in \cite{zhou_modeling_2022} to the varying sizes of attractions by changing the distance factor in the model to angular separation in \cite{xie_signage_2007}. In this way, the attention prediction is purely guided by ego-centric visual information, which maximizes its potential in generalizations. 
Also, Our model outperforms prior work in the prediction of attention terminations. While previous work by \cite{zhou_modeling_2022} chose a time-decaying function to represent how visual attentions fade, our study indicates such transitions can be better explained by pedestrian visual fields.

Then the locomotion model is conditioned on visual attention prediction. Compared to prior work that chose a fixed walking speed for pedestrians with visual attention \citep{weili_wang_modeling_2014}, we propose a dynamic walking speed scheme. We found that when pedestrians are attracted by environmental objects, their walking speed is regulated by the angular speed with regard to the objects. Such a setting can be related to experimental findings by \cite{warren_behavioral_2008}, where the author concluded that human locomotion is guided by optical flows.

Additionally, we show that our model is responsive to design feature variations, showing its potential in design optimizations. By configuring store widths and display depths, our model can be applied to various environmental objects such as stores, banners, and window displays. By adjusting corridor widths, architects can compare different design proposals with regard to facility dimensioning. Although the model in this paper is built upon a dataset of attention-based movement behaviors, its findings may give insights into similar topics such as signage detections or evacuation wayfinding.

Our work has some limitations that should be addressed in future work: First, in visual attention modelling, we only discuss a limited number of extrinsic factors in diluted pedestrian flows due to data availability. Future works should cover more extrinsic factors. Second, the empirical data we use is limited in the diversity of location and in pedestrian flow density. Our finding about factor influences on visual attention  does not apply to pedestrian flows with extreme high densities. Third, while attention-based movement behaviors play a dominant role in our context, integration of other response modes would be beneficial to create more realistic simulations. For example, the pedestrians who walk into the store. Fourth, although our visual attention model is compared to the original forms of prior models, a comparison that considers the varying parameter sizes of the models may better reveal the advantage of ours. Since we have not found a way to fairly compare models of which parameter size can not be easily scaled, we consider this topic as a potential future work. Finally, since the overall performance of our simulation framework is closely related to that of the Social Force Model, the limitations inherit in this model propagate, and such a base model can be improved to explore our attention-based model input.

\begin{table}
\caption{The final model fitting for visual attention modules. Parameters are fitted using standardized input variables (the means and the standard deviations are shown in the "Value" column). All parameter estimates were significantly different from zero (p-value $< 0.05$).}
\label{tab: sim_config}
\label{subtab:visual-attention}

\begin{tabular*}{\tblwidth}{@{\extracolsep{\fill}}lrlrrr@{}}
\toprule
Variable&&Parameter&    &    &    \\
Notation& Value &Notation&Estimate&Std. errors &Z scores  \\
\midrule
\textit{Attention initiation events} &    &    &    &    &    \\

 $\varphi$  & 0.981 $\pm$ 0.433 &$\theta_{0}$ &         3.167 &        0.351 &     9.031 \\
     $\phi$ & 1.797 $\pm$ 0.558 &$\theta_{1}$ &        -1.542 &        0.056 &   -27.308 \\
$\varphi^2$ & 1.151 $\pm$ 1.008 &$\theta_{2}$ &        -2.359 &        0.306 &    -7.709 \\
(intercept) & -                 &$\theta_{5}$ &        -4.683 &        0.107 &   -43.682 \\
\midrule
\textit{Attention termination events} &    &    &    &    &    \\

     $\varphi$ & 1.366 $\pm$ 0.383 &$\theta_{6}$ &        -0.804 &        0.160 &    -5.009 \\
        $\phi$ & 1.350 $\pm$ 0.504 &$\theta_{7}$ &        -2.510 &        0.363 &    -6.918 \\
      $\phi^2$ & 2.076 $\pm$ 1.472 &$\theta_{9}$ &         1.060 &        0.266 &     3.980 \\
 $\phi\varphi$ & 1.806 $\pm$ 0.776 &$\theta_{10}$&         0.828 &        0.193 &     4.284 \\
   (intercept) & -                 &$\theta_{11}$&         1.177 &        0.048 &    24.560 \\
\bottomrule
\end{tabular*}

\end{table}

\begin{table}
\caption{Attention-based locomotion regulation parameters in \S 3.3.3 that calculate desired speed.}
\begin{tabular*}{\tblwidth}{@{\extracolsep{\fill}}LLRL@{}}
\toprule
Parameter &           Description                & Value  & Unit \\
\midrule
$\meanidealangular$& Mean of ideal angular speed towards the store & 0.18 & rad/s\\ 
$\varianceidealangular$& Standard deviation of ideal angular speed towards the store & 0.04 & rad/s\\ 
\bottomrule
\end{tabular*}

\label{subtab:locomotion-regulation}
\end{table}

\begin{table}
\caption{Social Force Model parameters for low density scenarios, where asterisks denote the parameter values that differ from the original prior work \cite{zanlungo_microscopic_2012}.}
\begin{tabular*}{\tblwidth}{@{\extracolsep{\fill}}LLRL@{}}
\toprule
Parameter &           Description                & Value  & Unit \\
\midrule
$A$        & Weight of pedestrian repulsive effect & 1.13$^*$ & $m/s^2$\\
$B$        & Distance discounting factor of pedestrian repulsive effect & 1 & $m$\\
$r_v$      & Pedestrian perception range & 5.6 & $m$\\
$t_{max}$  & Maximum predicted collision time & 6.1 & $s$\\
$A^w $     & Weight of environment repulsive effect & 0.9 & $m/s^2$\\
$B^w $     & Distance discounting factor of environment repulsive effect & 1.0 & $m$\\
$\theta_v$ & Velocity bias & 0.16 & rad\\
$\lambda$  & Asymmetry parameter & 0.95 & \\
$\tau^{-1}$        & Inverse of relaxation time & 2$^*$ & $s^{-1}$\\
$\kappa$   & Weight of physical collision & 5000 \\
\bottomrule
\end{tabular*}
\label{subtab:sfm}
\end{table}

\begin{table}[]
\caption{All parameters in the boundary condition configuration.}

\begin{tabular*}{\tblwidth}{@{\extracolsep{\fill}}LLRL@{}}
\toprule
Parameter&           Description                  & Value & Unit\\
\midrule
$t_{+}$    & Mean time gap between pedestrians (direction 1) & 5.11 & s\\
$t_{-}$    & Mean time gap between pedestrians (direction 2) & 5.22 & s\\
$a_v$      & Parameter in neutral speed function & -0.00013 & \\
$b_v $     & Intercept of neutral speed function & 139 & cm/s\\
$\sigma_v$ & Standard deviation of desired walking speed&30&cm/s\\
$a_{\rho}$ & Distance regulation to the wall & 24.78 & cm \\
$b_{\rho}$ & Width factor of pedestrian flow & 0.2 & \\
$c_{\rho}$ & Positional factor of maximum density & 0.24 & \\
$d_{\rho}$ & Probability of walking on the wrong side & 0.36 & \\
$L_c$      & Width of the corridor & 540 & cm\\
\bottomrule
\end{tabular*}

\label{subtab:boundary}

\end{table}

\section*{Acknowledgements}

This material is based upon work supported by the U.S. Department of Homeland Security under Grant Award 22STESE00001-03-02. The views and conclusions contained in this document are those of the authors and should not be interpreted as necessarily representing
the official policies, either expressed or implied, of the U.S. Department of Homeland Security. The research was also supported in part by NSF IIS \#1955365, \#1955404, and NSF \#2324598.

Our study (ID: Pro2024000416) is identified as non-human subject research by Human Research Protection Program (HRPP) Institutional Review Board (IRB) at Rutgers University. The research activities do not meet the regulatory definition of human subjects research provided in 45 CFR 46.102. This project does not involve interaction or intervention with living individuals, nor does this project involve the use of identifiable private information or bio-specimens. 

The dataset used in the research originates from Tianchi Platform \\(https://tianchi.aliyun.com/dataset/162948).

\renewcommand*{\bibfont}{\small}
\bibliography{bibs}


\end{document}